\documentclass[useAMS,usenatbib,fleqn]{mn2e}
\usepackage{graphicx,color,pdfpages,amsmath,float}
\usepackage{subfigure,makecell}
\usepackage[table, dvipsnames]{xcolor}
\newcommand{\gadget}{\textsc{gadget-2}{ }}

\newcommand{\velociraptor}{\textsc{VELOCIraptor}{ }}
\newcommand{\treefrog}{\textsc{TreeFrog}{ }}
\newcommand{\orbweaver}{\textsc{OrbWeaver}{ }}

\def\ltsima{$\; \buildrel < \over \sim \;$}
\def\simlt{\lower.5ex\hbox{\ltsima}}
\def\gtsima{$\; \buildrel > \over \sim \;$}
\def\simgt{\lower.5ex\hbox{\gtsima}}
\def\la{$\langle$}

\title[Subhalo orbital trajectories]{Pre-processing, group accretion and the orbital trajectories of associated subhaloes}

\author[Bakels et al.] {\parbox{18cm}{
    Lucie Bakels$^{1,2,\star}$,
    Aaron D. Ludlow$^{1,2}$ and
    Chris Power$^{1,2}$
  }\vspace{0.3cm}\\
  $^{1}${International Centre for Radio Astronomy Research, University of Western Australia, 35 Stirling Highway, Crawley,}\\
  {Western Australia, 6009, Australia}\\
  $^{2}${ARC Centre of Excellence for All-Sky Astrophysics in 3D (ASTRO 3D)}\\
}

\begin{document}

\maketitle 

\begin{abstract}
  We use a high-resolution cosmological dark matter-only simulation to study the orbital trajectories of haloes and
  subhaloes in the environs of isolated hosts. We carefully tally all apsis points and use them to distinguish haloes
  that are infalling for the first time from those that occupy more evolved orbits. We find that roughly 21 per cent
  of resolved subhaloes within a host's virial radius are currently on first infall, and have not yet reached their first
  orbital pericentre; roughly 44 per cent are still approaching their first apocentre after infall. For the range of host
  masses studied, roughly half of all accreted systems were pre-processed prior to infall, and about 20 per cent were
  accreted in groups. We confirm that the entire population of accreted subhaloes -- often referred to as ``associated''
  subhaloes -- extends far beyond the virial radii of their hosts, with roughly half currently residing at distances
  that exceed $\approx 1.2\times r_{200}$. Many of these backsplash haloes have gained orbital energy since infall,
  and occupy extreme orbits that carry them well past their initial turnaround radii. Such extreme orbits are
  created during the initial accretion and dissolution of loosely bound groups, but also through penetrating
  encounters between subhaloes on subsequent orbits. The same processes may also give rise to unexpectedly abrupt
  {\em losses} of orbital energy. These effects combine, giving rise to a large variation in the ratio of sequent
  apocentres for accreted systems. We find that, within 2 virial radii from host centres, the concentrations of first-infall
  haloes are remarkably similar those of isolated field haloes, whereas backsplash haloes, as well as systems that were
  pre-processed, are considerably more concentrated. 
\end{abstract}

\renewcommand{\thefootnote}{\fnsymbol{footnote}}
\footnotetext[1]{E-mail: lucie.bakels@research.uwa.edu.au}

\section{Introduction}
\label{sec:intro}

In the standard cosmological model the variance of linear matter density fluctuations increases systematically
toward smaller scales. The dominant mass component is a cold and collisionless particle -- referred
to as cold dark matter (CDM hereafter) -- and the energy density of the Universe is dominated by dark energy
(denoted $\Lambda$ in the simplest case of a cosmological constant). This is the $\Lambda$CDM model, which has
few rivals in its ability to accurately describe the large- ($\simgt 10\,{\rm Mpc}$) and intermediate-scale
($1-10\,{\rm Mpc}$) structure of the Universe \citep[see][for a recent review]{FrenkWhite2012}.

$\Lambda$CDM also makes a few unwavering predictions. One, a consequence of the shape of the density fluctuation
power spectrum, is that structure formation proceeds hierarchically, from small to large scales.
Dark matter accretes onto primordial overdensities, either smoothly or through mergers, which grow
progressively more massive over time, forming gravitationally-bound dark matter haloes. The vestiges of a halo's past accretion
or merger events form a population of ``substructure'' haloes, or {\em subhaloes} for short. Haloes are the likely sites of
galaxy formation \citep[e.g.][]{White1978}, and subhaloes the potential hosts of satellite galaxies,
such as those observed around the Milky Way or other nearby galaxies, or of the individual galaxies in rich
clusters. The dynamics, spatial distribution and structure of substructure haloes therefore hold valuable clues to
the hierarchical nature of galaxy formation.

There is now a broad and comprehensive literature addressing a variety of issues related to the structure
and substructure of dark matter haloes. For example, it is now known
that substructure makes an important but sub-dominant contribution to the total mass of a halo, typically
$\simlt 10-15$ per cent \citep{Ghigna1998,Neto2007}, and that their mass function
is well-approximated by a single power-law, $d{\rm N}/d\log{\rm M}\propto {\rm M}^{-0.9}$
\citep[][]{Springel2001b,Gao2004,Springel2008b,Giocoli2010,Garrison-Kimmel2014}. Subhaloes are spatially
\citep[e.g.][]{DeLucia2004,Springel2008b,Gao2012} and kinematically \citep[e.g.][]{Gill2004,Diemand2004b,Sales2007a,Ludlow2009}
biased with respect to the underlying distribution of dark matter, which likely indicates
that different environmental process shape the orbital evolution of substructure compared to smoothly-accreted dark matter.

As with dark matter haloes, subhaloes can have substructure of their own, a hierarchy that can, in principle, extend
to the free-streaming limit of the dark matter particle (indeed, four nested levels of
subhaloes-in-subhaloes have been observed in highest-resolution simulation of the Aquarius Project; \citealt{Springel2008b}).
As a result, many subhaloes will have been be accreted as part of a substructure {\em group}
\citep[e.g.][]{LiHelmi2008}, or will have been pre-processed in some manner prior
to accretion \citep{Wetzel2015,Han2018,Bahe2019}.

In addition, when a halo or group of haloes is accreted by a more massive system, it will not necessarily
remain a subhalo indefinitely. Instead, accreted haloes can pass through temporary phases of
being nominal substructures -- i.e. confined to the virial boundaries of their host haloes -- and pass again into the field,
reaching orbital radii that can, in principle, extend to many times the host's virial radius
\citep[e.g.][]{Gill2005,Bahe2013,Haggar2020}. As a result, galaxies in the
field may show tell-tale signs of having passed through the dense central regions of more massive systems.
Indeed, simulations suggest that as many as {\em half} of all accreted systems currently lie at distances that exceed
the traditional virial boundaries of their hosts \citep{Ludlow2009}. These are often referred to as ``backsplash'', or
``associated'' subhaloes, terms we adopt in this paper.

This can have important consequences. For example, at fixed luminosity, galaxies hosted by backsplash haloes have higher mass-to-light
ratios \citep{Knebe2011b} and higher quenched fractions \citep{Simpson2018} than field galaxies.
The mass profiles of backsplash haloes are more concentrated than those of field haloes at comparable radial separations from their
hosts, at least partially explaining a puzzling phenomenon known as assembly bias
\citep{WangMo2009,Li2013,Sunayama2016,Mansfield2020}.

The accretion of groups of substructure expected in $\Lambda$CDM also has important consequences. One is that
accreted groups, when interacting with their host haloes near orbital pericentre, can lead to multi-body interactions,
a scenario favourable for the rapid exchange of orbital energy and angular momentum between group members. This can, in principle, 
result in the ejection of (typically low-mass) subhaloes on highly ``unorthodox'' orbits \citep{Sales2007b}, which often
propel them to distances that exceed their nominal turnaround radius \citep{Ludlow2009}. As we will see in
Section~\ref{ssec:prepdynamics}, the same interactions can also lead to an abrupt {\em loss} of orbital energy
and angular momentum, confining subhaloes to the innermost regions of their host haloes.

One consequence of group accretion, mentioned above, is that many systems, apparently infalling for the first time,
will have been ``pre-processed'' prior to accretion (i.e. will have been a subhalo of a more massive halo prior
infall).
Simulation work suggests that as many as half of the Milky Way's satellites (with ${\rm M_\star\simgt 10^6\,{\rm M_\odot}}$)
may have experienced such pre-processing \citep{Wetzel2015}, and likely more in galaxy groups and clusters.
\citet{Bahe2019}, for example, found that as many as 87 per cent of $\simgt 10^{10}\,{\rm M_\odot}$ haloes accreted by
massive clusters were pre-processed, and 73 per cent of Milky Way-mass ($\sim 10^{12}\,{\rm M_\odot}$) haloes were.
In group environments, these numbers are $\approx 70$ per cent and 35 per cent, respectively. 

There is also observational evidence for pre-processing. Group catalogues based on the Sloan Digital Sky Survey
(data-release 7; \citealt{Yang2007}) have been used to classify galaxies as infall or backsplash, central or satellite.
In the radial range between 2 to 3 virial radii from hosts, satellite galaxies show a higher quenched fraction than centrals, even when
controlling for backsplash galaxies \citep{Hou2014}. Results from the Local Cluster Substructure Survey (LoCuSS;
\citealt{Bianconi2018}) suggest that groups of galaxies infalling onto clusters possess lower star formation rates,
on average, than the total infalling population. These are only a couple of examples of the rich literature exposing
pre-processing as an important driver of galaxy evolution
\citep[see also, e.g.,][]{Cortese2006,Wetzel2014,Just2019}.

Our goal is to reassess some of these issues, aided by high-resolution cosmological (dark matter-only)
simulations. We focus our analysis on well-resolved and isolated primary haloes,
and consider {\em all} secondary haloes (whether subhalo or field halo) that ever come with
$4$ virial radii of their host. We carefully classify the trajectories of secondary haloes using sensible diagnostics
based on their orbital histories. Rather than drawing arbitrary distinctions between infalling and accreted
systems, we instead separate them based on the number of periapsis points, $N_{\rm peri}$, measured along their
trajectories: orbits for which $N_{\rm peri}=0$ are likely infalling for the first time; $N_{\rm peri}\geq 1$
represent more evolved orbits. As discussed below, many infalling haloes have been pre-processed,
a result that has important implications for not only their internal structure, but also for the ensuing evolution of
their orbital trajectories

Our paper is organized as follows. We describe our Numerical methods in Section~\ref{sec:methods}:
our simulations are described in Section~\ref{ssec:simulations}; our halo-finding techniques in
Section~\ref{ssec:velociraptor}; our merger trees and orbit tracking methods in Sections \ref{ssec:treefrog} and
\ref{ssec:orbits}, respectively. In Section~\ref{sec:results} we describe our main findings: Section~\ref{ssec:spatial}
focuses on the spatial distribution of associated and first-infall haloes and subhaloes; the distribution and evolution
of their apsis points are described in Sections~\ref{ssec:apsis} and ~\ref{ssec:apoevol}, respectively. The importance
of pre-processing and group infall relative to the accretion of ``pristine'' field haloes is discussed in
Section~\ref{ssec:preprocessing}, and the corresponding impact on subsequent dynamics in Section~\ref{ssec:prepdynamics}.
Finally, in Section~\ref{ssec:struct}, we comment on the importance of pre-processing and group-infall for the structural scaling
relations of dark matter haloes and subhaloes in the vicinity of massive systems. We summarize our results in
Section~\ref{sec:summary}.

\section{Simulations and Analysis}
\label{sec:methods}

\subsection{Simulations}
\label{ssec:simulations}
Our results are based on a high-resolution, cosmological dark-matter only simulation
carried out with a lean version of \gadget \citep{Springel2005b}. The simulation is
part of the {\small Genesis Simulations} suite of cosmological $N$-body runs
\citep[cf.][]{Poulton2019}, and follows the evolution of ${\rm N_{DM}}=2048^3$ collisionless
dark matter particles in a cubic box of comoving side-length $L=105\,h^{-1}{\rm Mpc}$. The softening
length, fixed in comoving units, is $\epsilon=1.7\, h^{-1}\,{\rm kpc}$ (roughly $1/30^{th}$ of the
Lagrangian mean inter-particle spacing); the particle mass is $m_p=1.17\times 10^7\,h^{-1}{\rm M_\odot}$.
Particle data are saved as snapshots at 190 discrete intervals between $z=20$ and $0$, equally
spaced in the natural logarithm of the expansion factor, which allows us to
robustly track the orbits and assembly histories of haloes and subhaloes. 

Initial conditions were created at $z=99$ by perturbing an initially uniform particle lattice
using second-order Lagrangian perturbation theory \citep{Scoccimarro1998,Crocce2006} in
accord with the linear power spectrum determined by the \citet{Planck2016}. The cosmological
density of baryons, matter and dark energy are
$(\Omega_{\rm bar},\Omega_{\rm M},\Omega_\Lambda)=(0.0491,0.3121,0.6879)$;
${\rm H_0}=67.5\,{\rm km/s/Mpc}$ is the Hubble-Lema\^{i}tre constant; $\sigma_8=0.815$ is the
linear rms density fluctuation in 8 Mpc spheres; $n_s=0.965$ is the spectral index of primordial power-law
density perturbations.

\subsection{Halo identification and selection}
\label{ssec:velociraptor}
Haloes and subhaloes are identified using \velociraptor \citep{Elahi2019a}, which operates
in two stages. First, distinct haloes are identified using a friends-of-friends (FOF) algorithm
\citep[e.g.][]{Davis1985} employing a linking length of $b=0.2$ times the mean inter-particle spacing.
Substructure haloes are then excised from each FOF group by searching for dynamically-distinct particle subsets in
six-dimensional phase-space (i.e. using a 6DFOF algorithm), and correspond to local overdensities whose velocity
distributions differ substantially from that of the locally-averaged background halo. 
Following \citet{vandenBosch2017} and \citet{vandenBoschJiang2016},
haloes and subhaloes resolved with fewer than 50 particles (at $z=0$) are
also discarded. Their centres coincide with the coordinate of the particle with the minimum
potential energy, and their bulk velocities are defined using the centre-of-mass velocity of all particles
in the 6DFOF envelope.

For FOF haloes, \velociraptor calculates various commonly-used definitions of virial mass.
For our analysis we adopt ${\rm M_{200}}$, the mass contained within a sphere of radius $r_{200}$
that encloses a mean density of $200\times \rho_{\rm crit}$ ($\rho_{\rm crit}=3 H^2/8\pi G$ is the
critical density for a closed universe). For {\em substructure}, masses are defined using the full
subset of particles that \velociraptor deems dynamically associated (i.e. the 6DFOF mass).
In addition to mass, we also calculate the radius ${\rm R_{max}}$ at which the circular
velocity profile reaches its maximum value, ${\rm V_{max}}$ -- we use these as non-parametric 
measures of the internal structure of haloes and subhaloes.

For our analysis, we identify a sample of 2309 primary haloes with virial masses
${\rm M_{200}}\ge 10^{12}\,h^{-1}{\rm M_\odot}$ (${\rm N_{200}}\simgt 8.6\times 10^4$) that also satisfy two
isolation criteria: one ensures that it is the most massive system within
eight times its virial radius, and the other that the host is not within eight
virial radii of a more massive system. We hereafter refer to these as {\em primary} haloes. The most
massive primary in our sample has a virial mass of ${\rm M_{200}}=3.36\times10^{14}\, h^{-1}{\rm M_\odot}$
(${\rm N_{200}}\approx 2.9\times 10^{7}$). 

Following custom, we refer to substructure haloes that lie within one virial radius $r_{200}$ of any
primary host as ``primary subhaloes''. Isolated ``field'' haloes
that lie beyond $r_{200}$ of primary hosts and {\em have never been} substructures of a more massive
system are referred to as {\em secondary} haloes (see Section~\ref{ssec:orbits} for details). 

\subsection{Merger trees}
\label{ssec:treefrog}

Halo and subhalo merger trees are constructed using \treefrog \citep[see][for details]{Elahi2018,Elahi2019b},
which is part of the \velociraptor software package. Moving forward (from high- to low-$z$) through the list of halo catalogues,
\treefrog links haloes identified in snapshot $S_i$ to their most probable descendant in snapshot $S_{i+1}$
by maximising a figure-of-merit, defined
\begin{equation}
  \Psi^2=\frac{{\rm N}^2_{S_i^k \cap S_{i+1}^l}}{{\rm N}_{S_i}^k\,{\rm N}_{S_{i+1}}^l}.
  \label{eq:fom}
\end{equation}
Here ${\rm N}_{S_i}^k$ is the number of particles in progenitor $k$ in snapshot $S_i$, ${\rm N}^l_{S_{i+1}}$ are
the numbers of particles in each unique descendant $l$ in snapshot $S_{i+1}$,
and ${\rm N}_{S_i^k \cap S_{i+1}^l}$ are the number of particles they have in common.

Rarely, idiosyncratic cases -- such as multiple mergers of approximately equal-mass systems -- lead to several
equally-probable descendants being identified. These issues are dealt with following \citet{Poole2017}, by
weighting each descendant's figure-of-merit by the rank-ordered binding energy of particles
$S^k \cap S_{i+1}^l$, where the rank ordering is carried out for particles in both the progenitor and descendant,
thereby maximizing the fraction of highly-bound mass between the two. Other
common failures of the algorithm arise when subhaloes temporarily disappear from the halo catalogues,
either during a close passage through the centre of a more massive system, or by dropping below \velociraptor's
20-particle detection limit. We mend such occurrences by searching for descendants across a series of
five consecutive snapshots until a suitable descendant is found; if one is not, we assume the halo to have
merged with its host, or to be tidally disrupted \citep[details of the algorithm can be found in][]{Poulton2019}.

One distinguishing feature of \velociraptor is its ability to identify dynamically cold
``substructure'' that is slightly overdense with respect to the {\em smooth}
background halo. A downside -- at least for our purposes -- is its tendency to identify subhaloes with strong tidal
features and tidal debris from recently {\em disrupted} subhaloes which must be distinguished from
gravitationally-bound substructure. These sorts of systems are transients in our halo catalogues
and can be easily identified by imposing constraints on their merger trees. For example, we find that the
vast majority are eliminated by demanding that each (sub)halo in the merger tree catalogue is able to be
tracked through at least 8 consecutive simulation snapshots (which corresponds to about one-quarter of the
circular orbital time at $r_{200}$, i.e. $t_{\rm orb}/4=(\pi/2)\, r_{200}/V_{200}$).

\subsection{Subhalo orbits and their classification}
\label{ssec:orbits}

\begin{figure}
  \centering
  \includegraphics[width=\linewidth]{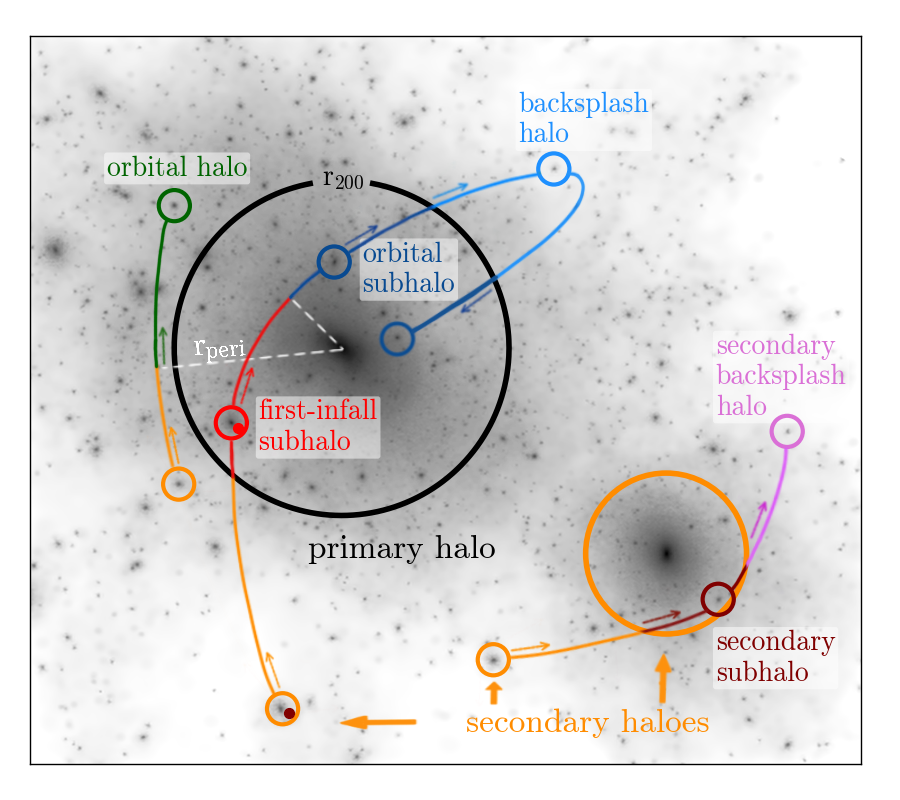}
  \caption{A visual representation of the various orbital categories defined in Section~\ref{ssec:orbits}.
    The solid black circle marks the virial radius, $r_{200}$, of a single primary host halo; isolated secondary haloes
    are shown using orange circles. Upon first crossing the virial radius of the primary, a secondary halo becomes a
    {\em first-infall} subhalo (red), and after crossing orbital pericentre, an orbital subhalo (dark blue). Orbital
    subhaloes whose trajectories carry them beyond the virial radius of the primary are labelled ``backsplash'' haloes
    (light blue) as long as $r_{\rm sub}\geq r_{200}$. Secondary haloes may also cross pericentre {\em outside} of
    their host's virial radius; these are ``orbital haloes'' (green). Similarly, secondary haloes may also have their
    own populations of secondary subhaloes (burgundy) or secondary backsplash haloes (magenta) -- these define a sample
    of ``pre-processed'' systems.} 
  \label{fig0}
\end{figure}

We study the dynamical and structural evolution of {\em surviving} (i.e. $z=0$) haloes and subhaloes that
have at some point in the past crossed within $4 \times r_{200}$ of any of the primary host haloes that
meet the isolation criteria described in Section~\ref{ssec:velociraptor}. Orbits are
calculated with respect to these haloes directly from the merger trees using \orbweaver \citep{Poulton2019}, which identifies any
apsis points along a halo or subhalo's trajectory about their primary host, as well as other quantities of interest
(e.g. orbital eccentricities, angular momenta, etc.).

The orbital histories of these (sub)haloes can be quite complex, and depend strongly on their interaction
history with the primary halo, or on interactions with other haloes in the primary's vicinity.
\citet{vandenBosch2017}, for example, identified 12 distinct evolutionary pathways through which subhaloes can evolve
between simulation outputs, including frequent penetrating subhalo-subhalo encounters and subhalo exchange between
distinct hosts.

Distinguishing haloes and subhaloes is non-trivial; one common approach is to differentiate them
  based on whether they happen to lie inside or outside their host halo's virial radius at a particular time, respectively.
  But as recently pointed out by \citet{Diemer2020}, this blurs the halo-subhalo distinction due to the somewhat arbitrary
  density thresholds used to define virial radii. These authors argue in favour adopting physically well-defined radii,
  such as the splashback radius \citep{DK2014}, to delineate haloes and subhaloes, but this raises additional concerns. Why,
  for example, should {\em any} infalling halo be reclassified as a subhalo simply due to its proximity to the host?

This motivates our approach, which is primarily to divide haloes and subhaloes into
distinct categories based on whether or not they have crossed orbital pericentre on their orbit about the primary
host halo, without appealing to arbitrary crossing radii, like $r_{200}$. Doing so results in two distinct groups, which we
now define:\\

\noindent {\bf Orbital haloes and subhaloes:} The full population of
haloes and subhaloes that have crossed pericentre at least once on an orbit defined relative to the
centre-of-mass motion of their primary. These include current subhaloes of the primary halo (i.e. those with $r_{\rm sub}\leq r_{200}$,
provided they have crossed pericentre), the entire population of ``backsplash'' haloes that were
once within $r_{200}$ but have since left, as well as a small fraction of haloes whose pericentres exceed
the virial radius of their primary host. Orbital haloes and subhaloes can be divided into three distinct subcategories:
\begin{enumerate}\setlength\itemsep{0.25em}

\item {\bf \em Orbital subhaloes} are currently within $r_{200}$ {\em and} have passed orbital pericentre
  at least once. Note that we do not restrict pericentres to radii $r_{\rm peri}<r_{200}$ and therefore a small fraction
  of orbital subhaloes, $\approx 2$ per cent, are infalling for the first time (i.e. their last orbital pericentre
  occurred at $r > r_{200}$, but their current separation is $r_{\rm sub}\leq r_{200}$).

\item {\bf \em Orbital haloes} have periapses that exceed $r_{200}$ and have never crossed 
  the virial radius of their host; they have never been classified as subhaloes. These haloes occupy
  grazing orbits, and are sometimes referred to as ``fly-bys''.
  
\item {\bf \em Backsplash haloes} are seemingly isolated systems that have
  crossed within the virial radius of their primary host at least once in the past, but have since
  left ($r_{\rm sub}\geq r_{200}$ at $z=0$).
\end{enumerate}

The second distinct group of haloes and subhaloes are characterized as follows:\\

\noindent {\bf First-infall haloes and subhaloes:} Any halo or subhalo, regardless of its radial
separation from the primary, that has not yet crossed periapsis on its orbit relative to the primary.
These include all haloes (and their subhaloes) at $r\ge r_{200}$ that are not classified as orbital (ii) or
backsplash (iii) haloes, as well as subhaloes of the primary (with $r_{\rm sub}\leq r_{200}$) that are on first approach
(i.e. ${\rm N_{peri}=0}$). We define the following distinct categories of first-infall systems:

\begin{enumerate}\setlength\itemsep{0.25em}\setcounter{enumi}{3}
\item {\bf \em Secondary haloes} are systems that have never been identified as a
substructure halo of a more-massive system and have not yet crossed pericentre on their orbit relative to the primary host.

\item {\bf \em Secondary subhaloes} are, at $z=0$, located within the virial radius of {\em any} halo outside
  the virial radius, $r_{200}$, of a primary host. In practice, $\approx 93$ per cent of these are subhaloes of {\em secondary haloes} (iv);
  $\approx 3$ per cent are subhaloes of backsplash haloes (iii), and 4 per cent are subhaloes of {\em secondary} backsplash haloes
  (vi, below).

\item {\bf \em Secondary backsplash haloes} were once secondary subhaloes, but have since veered beyond
  the virial boundary of their secondary host halo.

\item {\bf \em First-infall subhaloes} are within the virial boundary of their primary host, but
  have not yet crossed first pericentre.

\end{enumerate}

Figure~\ref{fig0} provides a visual representation of the various orbital types described above.
Note that secondary haloes (iv) and subhaloes (v) only exist beyond $r_{200}$ of their primary halo; after
crossing $r_{200}$ (but before reaching pericentre) we refer to them as first-infall subhaloes (vii) in order to
distinguish them from orbital subhaloes (i).
Secondary backsplash haloes (vi) and subhaloes (v) form a population of ``pre-processed'' systems that have
encountered dense environments beyond the virial boundary of the primary host. Of course, any
associated subhalo (a term we will often use for the combined population of orbital subhaloes (i),
backsplash haloes (iii) and first-infall subhaloes (vii)) may have been pre-processed as well, and
we will draw explicit distinctions between the two samples when necessary.

\begin{figure}
  \centering
  \includegraphics[width=\linewidth]{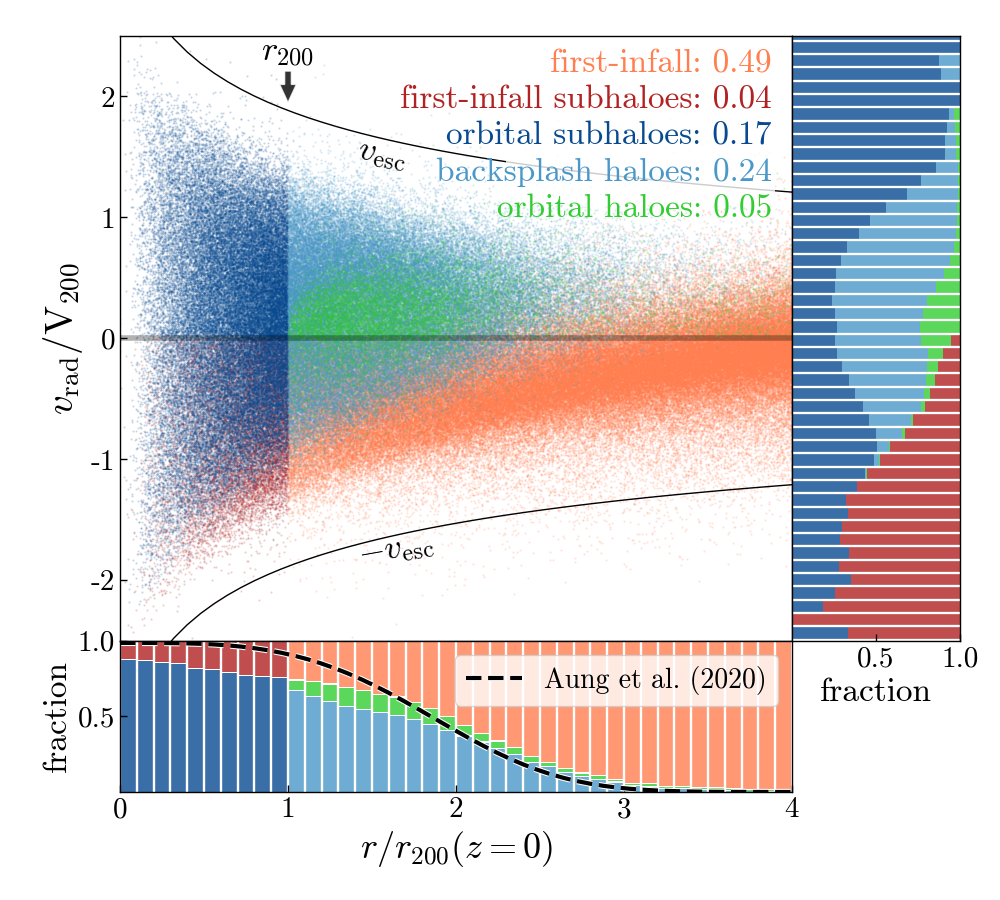}
  \caption{The main panel plots the phase-space diagram of radial velocity versus distance for haloes and
    subhaloes surrounding primary hosts. We combine all data for the 282 hosts that span the mass range
    $10^{13} \leq {\rm M_{200}}/[h^{-1}{\rm M_\odot}]\leq 10^{14}$ and also meet our isolation criteria (described in
    Section~\ref{ssec:velociraptor}). Coloured
    points differentiate the populations of (sub)haloes defined in Section~\ref{ssec:orbits}: orange
    and red points correspond to first-infall haloes and subhaloes, respectively; blue points
    to systems that have completed at least one pericentric passage at $r\leq r_{200}$, and are
    currently inside (dark) or outside (light) $r_{200}$. Green points are reserved for orbital
    haloes -- those that have completed one or more pericentric passages at radii exceeding $r_{200}$.
    The thin black lines demark the typical escape
    velocity curve for hosts in this mass range. The lower- and right-hand panels plot the
    fraction of (sub)haloes in each of these distinct samples that occupy discrete bins or radius
    or pair-wise radial velocity, respectively (for clarity, infalling haloes are left out of the latter). The
    black line in the bottom panel shows an estimate of the fraction of subhaloes that have completed an orbit
    about their hosts, as estimated by \citet{Aung2020}.}
  \label{fig1}
\end{figure}

\section{Results}
\label{sec:results}

\subsection{Spatial distribution of associated subhaloes and secondary haloes}
\label{ssec:spatial}

Figure~\ref{fig1} plots the phase-space diagram of radial velocity versus distance for orbital and 
first-infall haloes and subhaloes relative to their primary hosts. Data are shown for primaries in
the mass range $10^{13} \leq {\rm M_{200}}/[h^{-1}{\rm M}_\odot]\leq 10^{14}$, and use different colours to
highlight several examples of the orbital histories outlined above. Red and dark blue points, for
example, show ($r_{\rm sub}\leq r_{200}$) subhaloes of primary
hosts that are either on first-infall or have crossed first pericentre within $r_{200}$
of the primary, respectively. Light blue points correspond to primary backsplash haloes.
First-infall haloes and subhaloes that have not yet crosses $r_{200}$ of the
primary are plotted as orange points (for clarity, we have combined categories iv, v and vi).
A minority ($\approx 5$ per cent of all systems within $4\times r_{200}$) are orbital haloes, coloured green,
the majority of which ($\approx 94$ per cent) are beyond $r_{200}$ of the primary at $z=0$.

The lower panel shows the fraction of
haloes/subhaloes of each type as a function of distance from the primary. Not surprisingly,
orbital subhaloes (dark blue) dominate at radii $r_{\rm sub}\leq r_{200}$, making up roughly 78
per cent of classically-defined ``substructures'' (i.e. of all subhaloes with $r_{\rm sub}\leq r_{200}$).
First infall subhaloes (dark red) are nevertheless quite common,
accounting for $\approx 24$ per cent of subhaloes at $r_{\rm sub}\approx r_{200}$,  and of order 11 per cent
 at $r_{\rm sub}\simlt 0.1\times r_{200}$, where a central galaxy is expected
to dominate; approximately 22 per cent of {\em all} subhaloes within $r_{200}$ are on first infall. 
As previously noted \citep{Gill2004,Ludlow2009}, the population of backsplash haloes extends well beyond
$r_{200}$ of the primary: roughly half populate regions that exceed $1.5\times r_{200}$, and about $\approx 2$
per cent exceed $3\times r_{200}$. 

Another important point to note is that backsplash haloes (light blue points) tend to have a
slight positive radial velocity bias relative to their host, at least for the primary mass range
studied here. The mean radial velocity of {\em all} backsplash haloes, for example, is
$\approx 0.28\times {\rm V_{200}}$, which increases to $\approx 0.35\times {\rm V_{200}}$ for those at
radii $r_{\rm sub}\simgt 2\times r_{200}$. This suggests that many of these systems are still approaching
first their apocentre (after turnaround), and do not have ``backsplash'' radii (i.e. first apocentres)
that can be measured directly from their orbits \citep[see also][]{Diemer2017}. The right-hand panel
in Figure~\ref{fig1}, for example, shows the relative fractions of each sample in bins of radial
velocity.

\begin{figure}
  \centering
  \includegraphics[width=\linewidth]{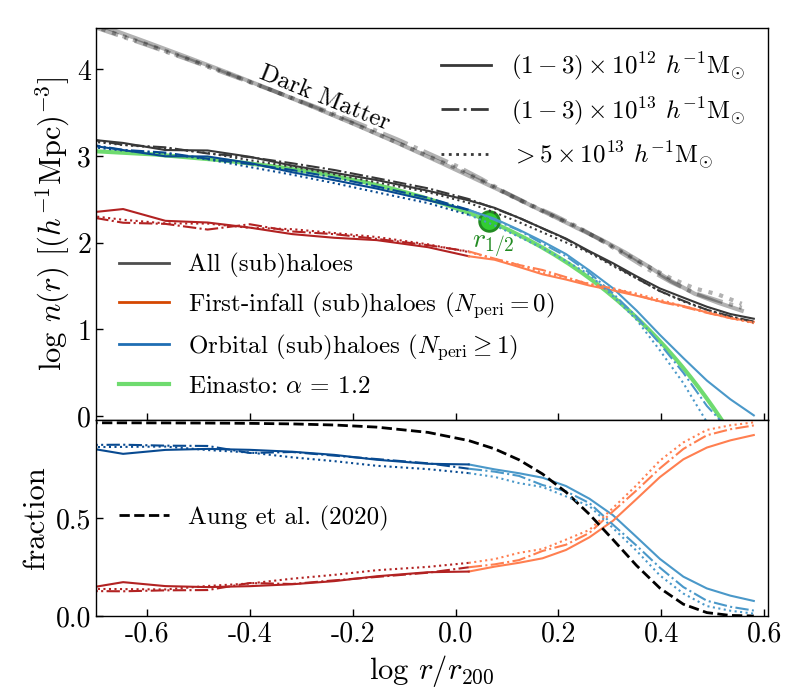}
  \caption{Spherically-averaged number density profiles of haloes and subhaloes surrounding primary hosts.
Results are are shown for three (primary) mass bins: $(1-3)\times 10^{12}\,{\rm M_\odot}$ (solid
lines; 1417 haloes), $(1-3)\times 10^{13}\,{\rm M_\odot}$ (dot-dashed; 211 haloes) and
$\geq 5\times 10^{13}\,{\rm M_\odot}$ (dotted; 37 haloes). Blue lines show the combined number
density profiles of orbital subhaloes, orbital haloes and backsplash haloes;
orange lines correspond to {\em all} first-infall haloes and subhaloes; black curves to
the entire sample. The green line shows the best Einasto-fit to all three associated subhalo
samples; the outsized circle marks the half-number radius of the fit, $r_{1/2}$, which exceeds $r_{200}$.
Note that each sample, independently or together, differs substantially from
the dark matter density distribution, which is shown using grey lines for each primary mass bin (normalised arbitrarily for comparison).
Note also the smoothness of the combined distributions: they do not exhibit distinct
features at $r_{200}$ that may be expected if that, or another arbitrary radius, were used as a
classifier of ``subhaloes''  or ``field haloes''. As in Figure~\ref{fig1}, we plot the fraction of
haloes and subhaloes in each sample as a function of radius in the lower panel, along with the
estimate of \citet{Aung2020} of the fraction of subhaloes that have completed at least one orbit.}
  \label{fig2}
\end{figure}

The number density profiles of orbital (blue curves) and first-infall haloes and subhaloes (red)
are plotted in Figure~\ref{fig2},
for three different bins of primary halo mass (the full populations are shown using black lines). 
The median spherically-averaged density profiles of dark matter (normalized arbitrarily for
comparison) are shown using grey lines. As hinted at in previous work
\citep[e.g.][]{Diemand2004b,Reed2005,Gao2004,Han2016}, the spatial distributions of the orbital haloes deviates
significantly from the dark matter, but can be approximated by an Einasto profile (\citealt{Einasto1965}; thick
green line). This density profile is given by
\begin{equation}
  \ln\biggr(\frac{n(r)}{n_{-2}}\biggl)=-\frac{2}{\alpha}\biggr(\biggr[\frac{r}{r_{-2}}\biggl]^\alpha-1\biggl),
  \label{eq:einasto}
\end{equation}
where $n_{-2}$ and $r_{-2}$ are characteristic values of density and radius where the logarithmic
slope, $d\ln n/d\ln r$, is equal to $-2$. The third parameter, $\alpha$, controls the shape of the
profile: $\alpha\approx 0.18$ yields one similar to the typical structure of dark matter haloes over the radial range
resolved by cosmological simulations \citep[e.g.][]{Navarro2010}; larger and
smaller values correspond to profiles that are more or less curved than that of the dark matter, respectively. 

Regardless of the mass of the primary, the spatial distribution of orbital haloes is well-approximated by eq.~\ref{eq:einasto}
with $\alpha\approx 1.2$ (shown as a thick green line in Figure~\ref{fig2}), and a concentration parameter
  $c\equiv r_{200}/r_{-2}\approx 2.6$ (the stacked dark matter haloes have considerably higher concentrations, between 4.8 and 7.7),
suggesting that they are
significantly less concentrated than the dark matter. Nevertheless, their number density profiles do not show
evidence of converging to a constant density core, but rather continue to rise all the way to the centre.


\subsection{Distribution of subhalo accretion times and apsis points}
\label{ssec:apsis}

\begin{figure*}
  \centering
  \includegraphics[width=0.9\linewidth]{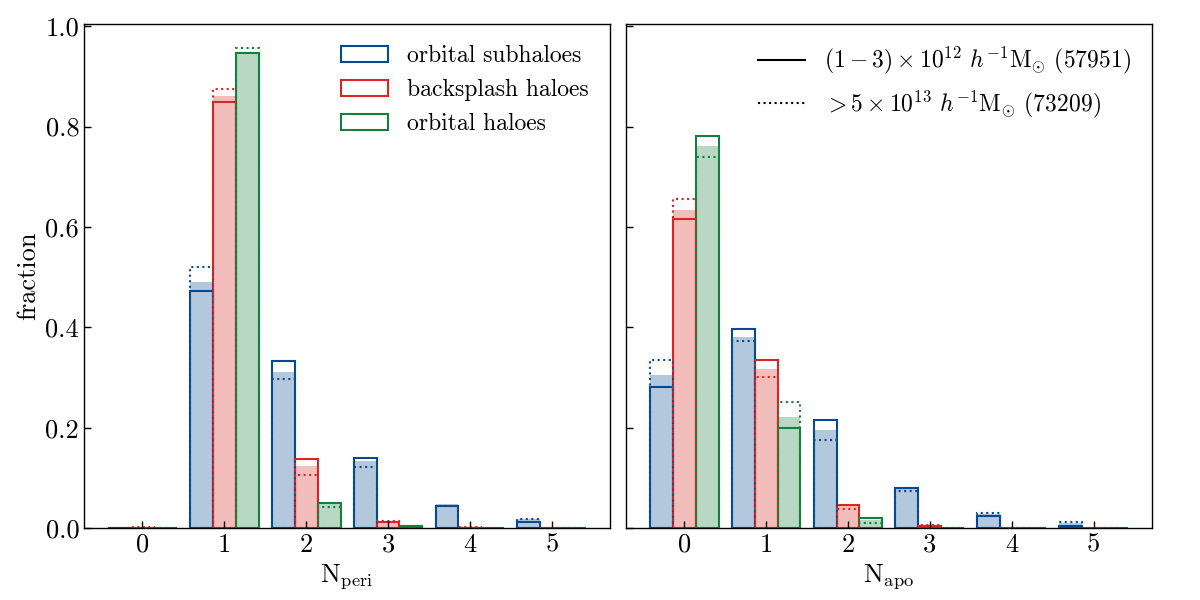}
  \caption{The fraction of orbital haloes and subhaloes, and backsplash haloes that have complete ${\rm N}$
    pericentres (left) and apocentres (right; measured after their initial turnaround radius). As with previous
    figures, different colours differentiate orbital
    type: blue corresponds to orbital subhaloes, red to backsplash haloes, and green to orbital
    haloes. Shaded bars show results after stacking all primary haloes, regardless
    of mass; solid and dotted bars show results when only hosts of mass 
    $(1-3)\times 10^{12}\,{\rm M_\odot}$ or $>5\times 10^{13}\,{\rm M_\odot}$ are included, respectively.} 
  \label{fig4}
\end{figure*}

In Figure~\ref{fig4} we plot the relative fraction of the number of apsis crossings for orbital
haloes (green), orbital subhaloes (blue) and backsplash haloes (red), showing separately 
pericentres (left) and apocentres (right; note that only apocentres occurring {\em after} the
initial turnaround radius are included). Shaded bars show results after combining
{\em all} primary haloes, but there is little variation across the mass range covered by our sample.
Solid and dotted open bars, for example, correspond to the lowest and highest host mass bins used
in Figure~\ref{fig2}.

As anticipated in Figure~\ref{fig1}, a large fraction of backsplash haloes, of order 63 per cent, are still
approaching their first orbital apocentre, and the vast majority, $\approx 86$ per cent, have crossed pericentre only once.
Only $\approx 5$ per cent have completed two or more apocentres. These numbers are even higher for orbital haloes:
76 per cent are still approaching first apocentre and $\simgt 95$ per cent have only one measured periapsis.
This suggests that backsplash haloes and orbital haloes are either relatively recent arrivals, or occupy
long-period orbits about their primary haloes. 
Intriguingly, many orbital {\em subhaloes} must have also been accreted relatively recently: at $z=0$ as many as 30
per cent have yet to reach first apocentre, and about half have crossed pericentre only once. Less than
$\approx 21$ per cent of orbital subhaloes have crossed pericentre three or more times; less than $\approx 13$
per cent have three or more measured apocentres. A significant fraction of orbital haloes, subhaloes and
backsplash haloes are therefore unlikely to be well-mixed with the potential of their primary host haloes.

\begin{figure}
  \centering
  \includegraphics[width=\linewidth]{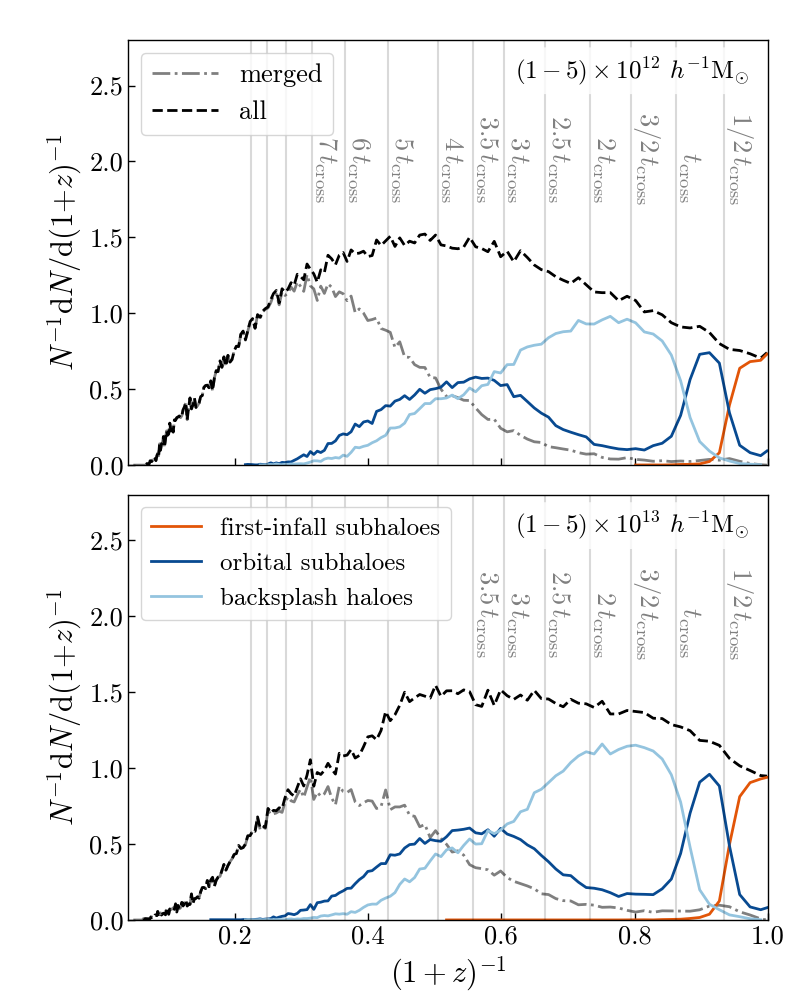}
  \caption{Distribution of subhalo accretion times for hosts in two separate mass bins (upper and lower panels).
    The red lines corresponds to first-infall subhaloes (i.e. ${\rm N_{peri}=0}$ and $r_{\rm sub}\leq r_{200}$ at $z=0$);
    dark blue lines to orbital subhaloes (${\rm N_{peri}\geq 1}$ and $r_{\rm sub}\leq r_{200}$ at $z=0$); light blue lines
    to backsplash haloes (${\rm N_{peri}\geq 1}$ and $r_{\rm sub}> r_{200}$ at $z=0$). Dot-dashed
    grey lines show the distribution of accretion times for subhaloes that, by $z=0$, have either merged
    with their host haloes, were tidally disrupted, or fell below our 50-particle limit.
    The black dashed lines correspond to the entire subhalo population. Thin vertical lines mark
    integer multiples of half-crossing times, defined $t_{\rm cross}=2\,r_{200}/{\rm V_{200}}$.}
  \label{fig:zacc}
\end{figure}

We explore this further in Figure~\ref{fig:zacc}, where we plot the distribution of subhalo accretion times for two separate
bins of host mass. As expected, essentially all first-infall subhaloes (red line) were accreted recently. The distribution
peaks at $z=0$, and is largely confined to infall times that do not exceed a half-crossing time\footnote{A crossing, $t_{\rm cross}=2\,r_{200}/V_{200}$,
  time is roughly the time required for recently accreted material to reach orbital pericentre.}, defined
$t_{\rm cross}/2=r_{200}/{\rm V_{200}}$ (shown as the right-most vertical grey line in Figure~\ref{fig:zacc}).
Interestingly, orbital subhaloes (dark blue lines) have a strongly bimodal accretion time distribution: one peak clearly confined
to infall times spanning $t_{\rm cross}/2\leq t_{\rm acc}\leq t_{\rm cross}$ (these are primarily subhaloes with
${\rm N_{apo}}=0$) and another peaking at
$t_{\rm acc}\approx 3.5 \times t_{\rm cross}$; the minimum between the two is reached at roughly
$t_{\rm acc}\approx 1.5\times t_{\rm cross}$. Not surprisingly, the minimum in the accretion time
distribution for orbital subhaloes coincides with a {\em maximum} for backsplash haloes (light blue lines), since accreted
material will typically require at least $t_{\rm cross}\approx 1.8 \,{\rm Gyr}$ to exit the virial boundary
of a halo once accreted. Indeed, the vast majority of backsplash haloes were accreted by their hosts at lookback
times exceeding this timescale (second vertical grey line from the right).

For comparison, we also plot the distribution of infall times for ``merged'' subhaloes, i.e. those that
have merged with their host, were tidally disrupted or fell below our imposed 50-particle limit. The majority of
these systems were accreted well before those that survive to the present day. For example, the peak
accretion rate of merged subhaloes occurred around $z\approx 2.3$. Averaged over all hosts, we find that 95 per cent
of subhaloes that {\em survive} to $z=0$ were accreted since $z=1.37$, whereas 95 per cent of those
that {\em do not}, have accretion redshifts $z_{\rm acc}\simgt 1.91$.

\begin{figure}
  \centering
  \includegraphics[width=\linewidth]{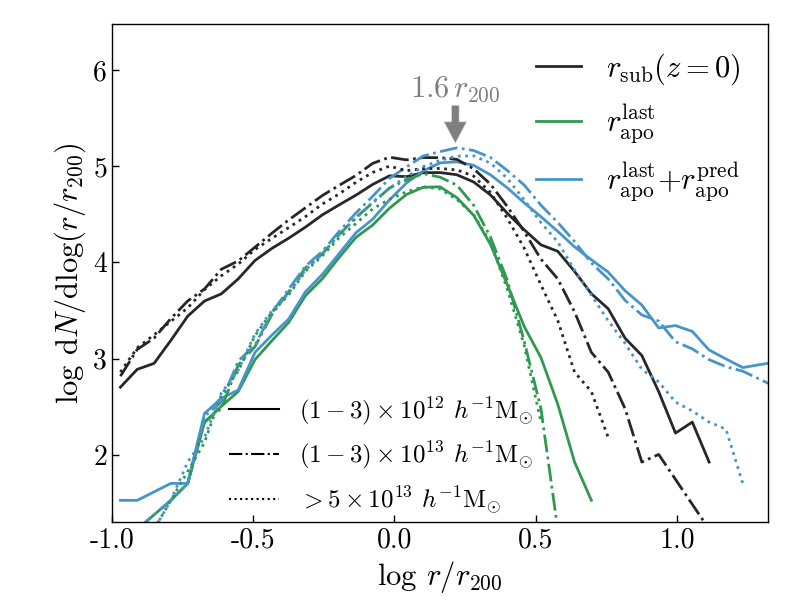}
  \caption{Distributions of associated subhalo radii, $r_{\rm sub}(z=0)$ (black curves), last-measured 
    apocentres, $r_{\rm apo}^{\rm last}$ (green lines), and the combined distribution of $r_{\rm apo}^{\rm last}$
    and $r_{\rm apo}^{\rm pred}$ (blue lines), where the latter are the {\em predicted} apocentres for systems that have
    not yet had one (the predictions are based on the radial energy equation; see text for details). Results are shown for the
    same three mass bins used in previous figures: $(1-3)\times 10^{12}\,{\rm M_\odot}$ (solid), $(1-3)\times 10^{13}\,{\rm M_\odot}$
    (dot-dashed) and $\geq 5\times 10^{13}\,{\rm M_\odot}$ (dotted). Note that the distribution of last-measured apocentres
    is a poor indicator of the true apocentre distribution; this is primarily a result of the large fraction of
    associated subhaloes, around 54 per cent, that are (at $z=0$) still approaching their first apocentre after turnaround.}
  \label{fig5}
\end{figure}

Overall, these results suggest that robustly measuring of the distribution of orbital
apocentres directly from the orbital tracks of associated subhaloes may be challenging, since many
have not yet had one. We demonstrate this in
Figure~\ref{fig5}, where we compare the current radial distribution of all associated subhaloes
(black curves) to the distribution of their {\em last measured} apocentres ($r_{\rm apo}^{\rm last}$;
green lines). As noted above only $\approx 46$ per cent of associated subhaloes have 
at least one measured apocentre, and the majority of those
that do not are backsplash haloes currently residing at radii $r\simgt r_{200}$.
This explains why the radial distribution of associated subhaloes is broader than the distribution of their last-measured
apocentres: haloes at very small radii are unlikely near apocentre, and so $r_{\rm sub}\simlt r_{\rm apo}^{\rm last}$,
whereas many of those at large radii have not yet reached apocentre, and are not included in the plot.
As we discuss below, this may have important implications
for measuring the so-called ``backsplash'' radius of primary haloes by appealing to measurements
of the apocentres of recently accreted material. This is an important caveat to consider, particularly
if backsplash radii are to be used as observational probes of cosmology \citep[e.g.][]{Adhikari2018}.

Nevertheless, it is possible to improve matters by {\em estimating} the orbital apocentres for
associated subhaloes when one cannot be directly measured. Knowing the spherically-averaged potential of
the host halo, $r_{\rm apo}$ (occurring at $z\leq 0$ in these cases) can be estimated directly\footnote{We have verified
  the robustness of this procedure by comparing the last-measured apocentre to the predicted one for a
  sample of associated subhaloes with ${\rm N_{\rm apo}}\geq 1$; the errors on the predicted $r_{\rm apo}$
  are typically random, and $\simlt 10$ per cent. } from the subhalo's orbital
kinetic energy and angular momentum (see eq. 3.14 of \citealt{BT2008}). Adding these predicted apocentres to
the last-measured ones yields an apocentre distribution (blue lines) that peaks at $\approx 1.6\times r_{200}$ (downward
arrow), which coincides with the value for periodic orbits anticipated by the self-similar infall model of
\citet{Bertschinger1985}. This is roughly 23 per cent larger than the peak of last-measured
apocentres, occurring at $\approx 1.3\times r_{200}$. Note also the extended tail
of $r_{\rm apo}$: depending on primary mass, between roughly 10 to 13 per cent of associated subhaloes
have $r_{\rm apo}\simgt 3\times r_{200}$,
with between 1 to 4 per cent exceeding $5\times r_{200}$. As discussed by \citet{Ludlow2009}, many of these
haloes ($\approx 37$ per cent in their study) occupy orbits that carry them beyond their initial
turnaround radius, a scenario difficult to reconcile with simple analytic infall models. 

\begin{figure*}
  \centering
  \includegraphics[width=\linewidth]{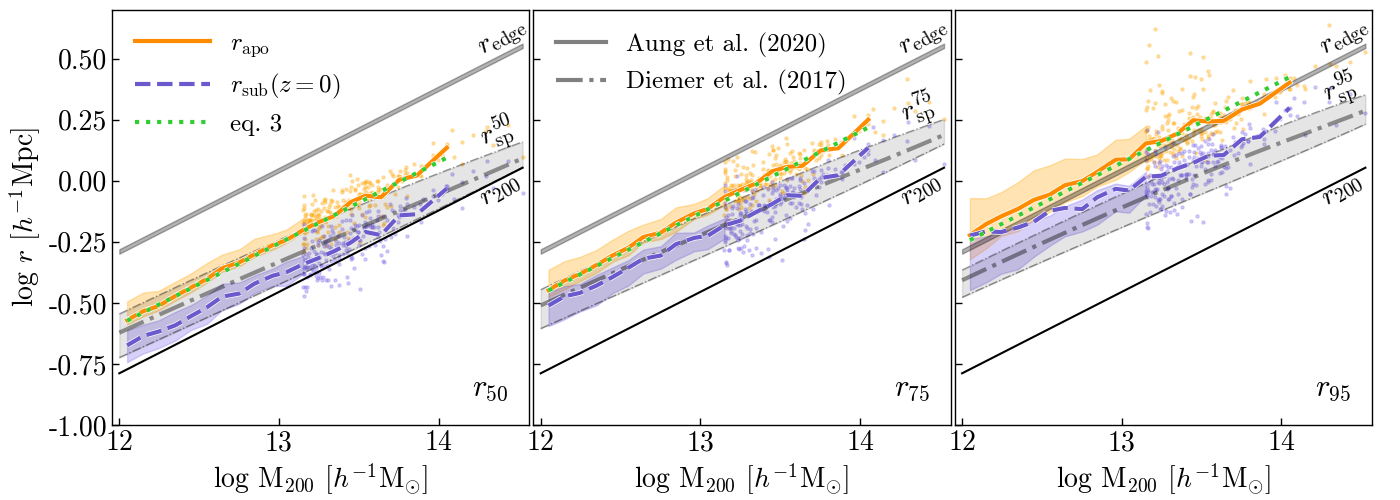}
  \caption{Radii that enclose 50, 75, and 95 per cent (left to right, respectively) of all associated subhaloes
    (blue dashed lines) and their apocentres (solid orange line) as a function of the virial mass of their primary. 
    Lines correspond to the median trends measured after stacking all hosts in equally-spaced logarithmic mass bins
    of width $\Delta \log {\rm M_{200}}= 0.1$; the shaded regions indicate the $25^{th}$ and $75^{th}$ percentile scatter for
    masses below $1.4\times10^{13}\,h^{-1}{\rm M_\odot}$. Above this mass scale, where there are fewer than 40 haloes per bin,
    results for individual haloes are shown using coloured points. For
    comparison, we also show the radii (and corresponding scatter) enclosing the same fraction $f$ of dark matter particle
    splashback radii as thick grey lines (labelled $r^f_{\rm sp}$ in each panel), which were estimated using the empirical fit
    provided by \citet{Diemer2017b}. The thick grey line marked ``$r_{\rm edge}$'' in each panel was determined by
    \citet{Aung2020} to enclose 99 per cent of associated subhaloes.}
  \label{apomax}
\end{figure*}

We characterize the mass-dependence of the spatial distribution and apocentres of associated subhaloes
quantitatively in Figure~\ref{apomax}. From left to right, different panels correspond to the ($z=0$) radii
enclosing 50, 75 and 95 per cent of all associated subhaloes (dashed purple lines), and their apocentres
(solid orange line). For comparison, the solid black lines indicate $r_{200}$ and thick solid grey lines
$r_{\rm edge}=1.96\times r_{200, {\rm mean}}$, determined by \citet{Aung2020} to enclose roughly 99 per cent
of associated subhaloes\footnote{Note that $r_{200,{\rm mean}}$ is the radius that encloses a mean 
  density of $\Omega_{\rm M}\rho_{\rm crit}$, which is slightly larger than $r_{200}$. For the purposes of
  Figure~\ref{apomax}, we assume an NFW profile when converting $r_{200,{\rm mean}}$ to $r_{200}$, and the
  thickness of the line indicates the variation expected for concentration parameters spanning $c=5$ to 15.}
(note that these authors used the radial velocity distributions of haloes and subhaloes to distinguish
infalling from accreted systems rather than explicitly tracking the orbits of individual objects). Thick
dot-dashed grey lines enclose the equivalent fraction of splashback
radii for dark matter particles (defined as the first apocentre after infall) determined from the fitting formulae of
\citet{Diemer2017b}.

These measures of the spatial distribution of associated subhaloes differ in detail, but
all paint a consistent picture: the spatial distribution extends far beyond $r_{200}$, and many associated
subhaloes occupy orbits with unexpectedly large apocentres. For example, we find that only about half 
have apocentres that confine them to within $1.6\times r_{200}$, and roughly 5 per cent
have apocentres that exceed $3.4\times r_{200}$. The radius enclosing a fraction $f_{\rm apo}$ of all associated
apocentres can be accurately described by the following fitting formula:
\begin{equation}
  \label{eq:fitapo}
  \log r(f_{\rm apo}) = \log r_{200} + \,a \log{f_{\rm apo}} + (1-f_{\rm apo})^b - c\,
\end{equation}
where $a = 0.321$, $b = -0.089$ and $c = 0.767$. Eq.~\ref{eq:fitapo}, shown as a dotted
green line in each panel of Figure~\ref{apomax}, is valid for $0.05 \simlt f_{\rm apo}\simlt  0.98$, and for host
masses between $10^{12}\,{\rm M_\odot}$ and $10^{14}\,{\rm M_\odot}$. 

Such extreme apocentres are not reached on their own, but require a considerable injection of orbital energy, likely
through 3-body encounters with other associated subhaloes, or as a result of a group accretion event. We will return
to this discussion in Section~\ref{ssec:prepdynamics}, but first turn our attention to the evolution of successive
orbital apocentres.

\subsection{The evolution of orbital apocentres}
\label{ssec:apoevol}

\begin{figure*}
  \centering
  \includegraphics[width=0.9\linewidth]{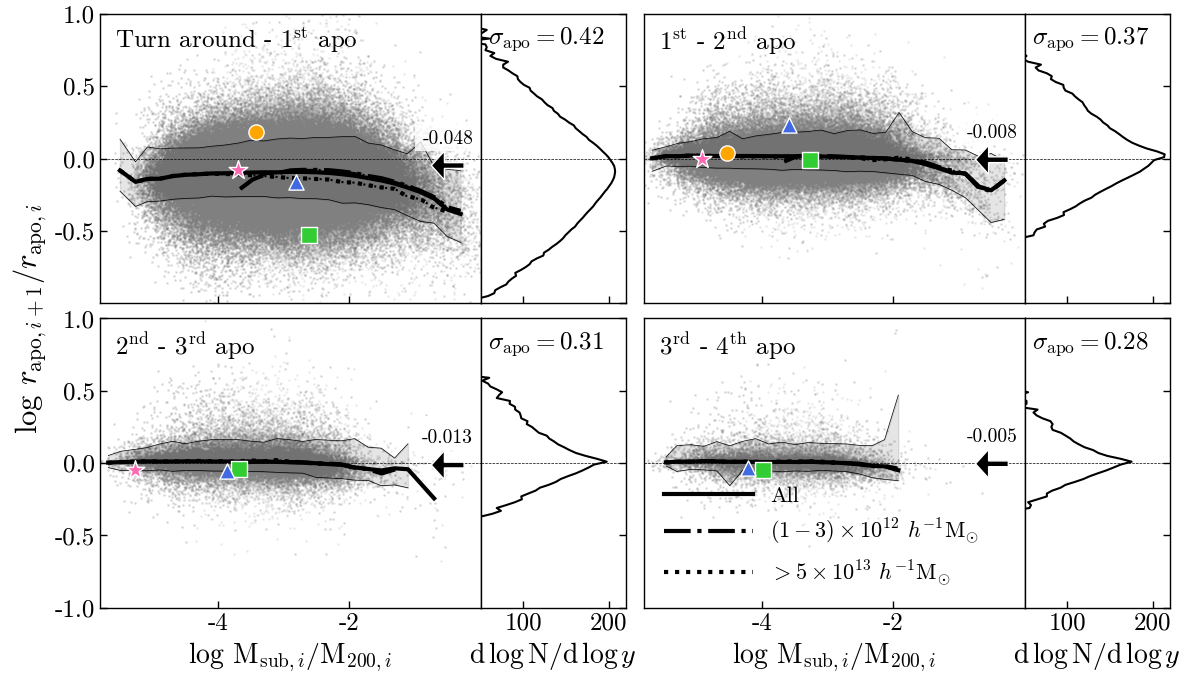}
  \caption{Ratio of successive apocentres for associated subhaloes plotted
    as a function of subhalo mass (expressed in units of the mass of the primary at apocentre $i$). From top-to-bottom, left-to-right
    panels correspond to the ratio of the first apocentre to the turnaround radius, $r_{\rm ta}$,
    then to the ratios of second-to-first, third-to-second, and fourth-to-third apocentres. In all panels,
    the mass of the subhalo and primary are measured at the first relevant apocentre (e.g. at the turnaround radius
    for those plotted in the upper-left panel). Note that only a small fraction, $\simlt 1$ per cent, of all accreted haloes  
    have $>4$ measurable apocentres and have been left out of the plot to avoid clutter. Individual (sub)haloes are shown using
    grey points, regardless of the mass of the primary; the thick black line and grey shaded region highlight the corresponding
    median and 25/75 percentile scatter. Medians for two separate primary mass bins, $(1-3)\times 10^{12}\,{\rm M_\odot}$
    (dot-dashed lines) and $\geq 5\times 10^{13}\,{\rm M_\odot}$ (dotted lines), are also shown.
    The distributions are shown to the right ($\sigma_{\rm apo}$ in each side panel indicates the linear variance);
    the black arrows mark the values of $r_{{\rm apo},i+1}/r_{{\rm apo},i}$ anticipated by Bertschinger's (1985)
    spherically-symmetric self-similar infall model.
    Note that orbital energy loss is most pronounced between turnaround and first
    apocentre after accretion, and exhibits a relatively strong mass dependence for
    ${\rm M_{sub}}/{\rm M_{200}}\simgt 0.01$. Outsized coloured points mark specific
    subhaloes whose trajectories are plotted in Figure~\ref{apoevolex}.}
  \label{apoevol}
\end{figure*}

Figure~\ref{apoevol} plots the ratio of successive apocentres, $r_{{\rm apo},i+1}/r_{{\rm apo},i}$, versus the
subhalo-to-primary mass ratio at the initial one, $i$. Dots correspond to individual (associated) subhaloes;
different lines mark the median relations for the entire sample of primaries (solid line), and for the highest (dotted;
$\simgt 5\times 10^{13}\,h^{-1}{\rm M_\odot}$) and lowest mass ones (dot-dashed; $\simgt (1-3)\times 10^{12}\,h^{-1}{\rm M_\odot}$).
The shaded regions show the $25^{th}$ and $75^{th}$ percentile scatter for the entire sample. Different
panels correspond to more or less evolved orbits: the ratio of first apocentre to turnaround radius
is shown in the top left panel, for example, and fourth-to-third apocentre ratio in the bottom-right.
Dotted horizontal lines correspond to successive equal-amplitude apocentres. Subhaloes lying above
or below these lines must have either gained or lost orbital energy along that segment of their orbit, respectively.

A few results are worth emphasizing. First, note  that 
the majority of orbital energy loss occurs after {\em first} pericentric passage, i.e. between
turn around and first orbital apocentre (top-left panel). On average, the ratio of first apocentre to
the turnaround radius is $\approx 0.8$. Subsequent apocentres, at least for {\em low} mass
subhaloes, tend to have comparable amplitudes, on average. This is indeed the trend expected from
self-similar infall models, in which accreted material gradually loses energy, eventually reaching
a stable orbit whose successive apocentres have equal amplitude. The arrows in each panel of
Figure~\ref{apoevol}, for example, show the ratios
anticipated by Bertschinger's (1985) model, which describes our simulation results reasonably
well\footnote{We note that in the Bertschinger model the first apocentre of an accreted shell
  is of order 90 per cent of its turnaround radius, slightly larger than the results for the subhaloes
  plotted in the upper-left panel of Figure~\ref{apoevol}.}.

The effect, however, discussed in detail in \cite{vandenBosch2016}, is mass dependent:
subhaloes with masses exceeding $\simgt 0.01\times {\rm M_{200}}$, for example, tend to experience an increased
drain of orbital energy due to the increasing importance of dynamical friction; lower mass subhaloes
behave as test particles in the potential of the host, and the mass-dependence of apocentre ratios
weakens substantially. This effect ultimately leads to a spatial segregation of subhaloes according to
their mass at infall, with more massive systems congregating toward the centres of haloes \citep[see, e.g.][]{Nagai2005,Ludlow2009,vandenBosch2016}.

Note also the considerable scatter along the ordinate axes of Figure~\ref{apoevol} (the rms variance,
$\sigma_{\rm apo}$, is quoted in each panel). Indeed, as many as 11 (5.3) per cent of associated subhaloes have
{\em first} apocentres that exceed their turnaround radius by as much as 25 (50) per cent, indicating a significant gain
in orbital energy after infall. Subhaloes may also {\em lose} energy after infall, and not always gradually, as
would be expected from, e.g., dynamical friction or self-similar infall models.
Roughly 42 (13) per cent have first apocentres
that are at least 25 (50) per cent {\em smaller} than $r_{\rm ta}$
(recall that, in Bertschinger's model, the first apocentre of an accreted shell has an apocentric
distance $r_{\rm apo}\approx 0.9\times r_{\rm ta}$).
Although virtually all of the most massive accreted systems
(${\rm M_{sub}}\simgt 0.1\times {\rm M_{200}}$) experience a substantial reduction in orbital energy
after infall, many {\em low-mass} systems -- presumably unaffected by dynamical friction -- do too.
This unexpected result requires explanation. We will return to this point in Section~\ref{ssec:prepdynamics}.

\begin{figure}
  \centering
  \includegraphics[width=\linewidth]{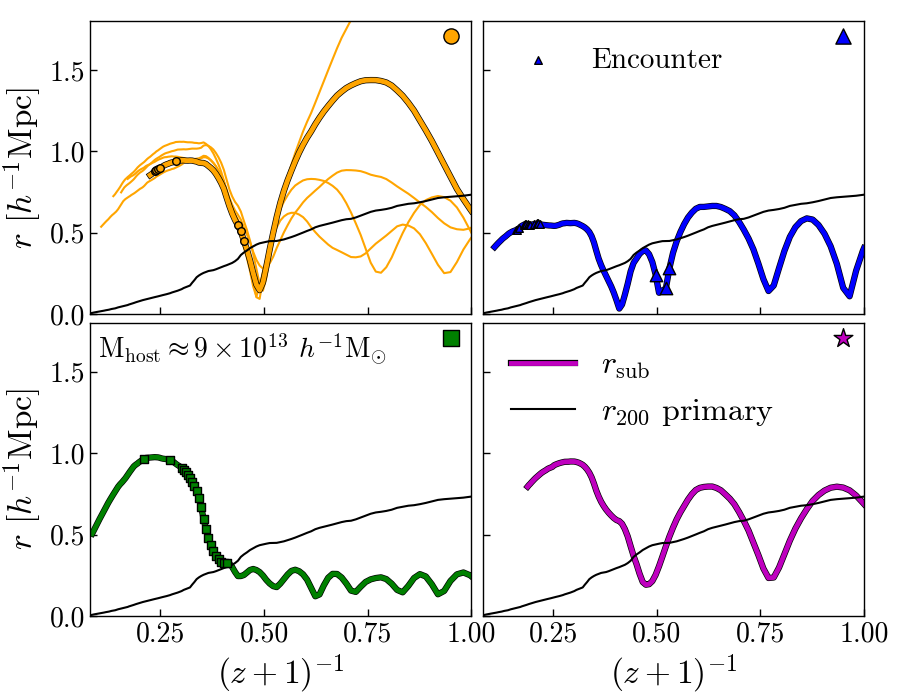}
  \caption{Orbital trajectories for the individual haloes and subhaloes highlighted in Figure~\ref{apoevol}.
    Identifying symbols, shown in the upper-right corner of each panel, are chosen to match those used in
    Figure~\ref{apoevol}. The upper-left panel shows an example of an extreme orbit (thick yellow line). This
    particular subhalo was accreted as part of a group (thin yellow lines correspond to orbits of other group members
    identified at the time of accretion) and was propelled onto a highly unorthodox orbit during the group's
    tidal dissociation near first pericentric passage. The dissociation of accreted groups not only ejects subhaloes
    along highly eccentric orbits, but can also lead to a rapid loss of orbital energy and angular momentum,
    resulting in some subhaloes becoming trapped on highly bound orbits, an example of which is shown
    in the lower-left panel (note that snapshots in which these infalling haloes were identified as secondary
    subhaloes -- i.e. subhaloes of an infalling secondary halo -- are marked using symbols). Penetrating encounters
    between subhaloes, particularly near orbital pericentre,
    can also lead to energy exchange (upper-right panel). For comparison, we plot a ``typical'' orbital trajectory
    in the lower-right panel. The virial radius $r_{200}(z)$ of the primary halo is shown using a black line in
    all panels.}
  \label{apoevolex}
\end{figure}

The outsized coloured points in Figure~\ref{apoevol} show a few examples of
subhaloes that illustrate these points; their orbits are shown in Figure~\ref{apoevolex}. In each case,
the loss or gain of orbital energy is invariably associated with an interaction. The upper-left panel, for
example (marked as a yellow circle in Figure~\ref{apoevol}),
shows a classic example of an ``ejected'' subhalo. Such extreme orbits are generated when accreted groups are
tidally-dissolved by the main halo near their orbital pericentre, redistributing their orbital energy and often propelling
low-mass members onto highly eccentric orbits. The remaining thin yellow lines show the orbits of other group members, which
exhibit a broad range of first apocentres, from $r_{\rm apo}\approx 0.6\, r_{\rm ta}$ to $r_{\rm apo}\approx 2.6\, r_{\rm ta}$.
At least two members of this group are placed on orbits so extreme that they are effectively removed from the
system as a whole. 

Of course, the opposite is also possible and energy can be abruptly {\em lost} during the tidal disruption of groups,
and one such example is evident in the upper-left panel of Figure~\ref{apoevolex}. Another, more extreme case is shown
in the lower-left panel (green square in Figure~\ref{apoevol}). This subhalo is also a member of
an accreted group (symbols along the orbital trajectories mark snapshots where each infalling halo was identified as a
secondary subhalo of a more massive secondary halo), whose tidal disruption led to a rapid reduction in orbital energy;
indeed, its first apocentre after infall is only of order $0.3\, r_{\rm ta}$, and remains comparable thereafter. 

While the types of interactions leading to abrupt changes in subhalo orbits 
are most common among members of infalling groups, the high number densities of subhaloes within
$r_{200}$ means that penetrating encounters\footnote{To estimate the frequency of penetrating
  encounters we first estimate the ``size'' of a subhalo as the virial radius $r'_{200}$ corresponding to its 6D FOF mass.
  Encounters only occur after first infall, and are identified by tagging all subhaloes that cross another's $r'_{200}$.}
between them, which may also perturb their orbits, should be relatively common
\citep[see e.g.][]{Tormen1998b,vandenBosch2017}. This, in principle, allows for
less likely scenarios, such as the one plotted in upper-right panel of Figure~\ref{apoevolex}. Here, the accreted halo
loses orbital energy on first pericentre ($r_{\rm apo,1}/r_{\rm ta}\approx 0.7$) but regains even more on the second
and is pushed to a third apocentre that exceeds it original turnaround radius ($r_{\rm apo,2}/r_{\rm ta}\approx 1.2$). 
Interestingly, this subhalo experienced a penetrating encounter with another close to its second orbital
pericentre (triangles along the orbital trajectory indicate when this encounter occurred), leading to a 3-body encounter
(the primary is the third body) favourable to the exchange of orbital energy and angular momentum.

It is also worth mentioning that instances of group accretion and penetrating encounters are not the only means
of provoking unorthodox orbits such as these. In fact, any rapid change in the gravitational potential of the host -- which
may be brought about by a merger, or any other large-scale fluctuation in the gravitational potential -- can also
lead to unexpected changes in the orbits of subhaloes as violent relaxation drives the system towards a new equilibrium. 

While such unorthodox orbits are not difficult to identify, they are aberrant. Indeed, many subhaloes
follow well-defined orbits that are not too dissimilar from those anticipated by simple spherical infall models.
The purple curve in the lower-right panel (purple stars in Figure~\ref{apoevolex}) shows a typical example of how subhalo
orbits evolve. We will see below that the outliers in the distributions of apocentre ratios are mainly associated with systems
that have been ``pre-processed'' by groups prior to infalling onto the primary halo.

\subsection{Pre-processing and the importance of group accretion}
\label{ssec:preprocessing}

A number of previous studies have identified pre-processing as an important regulator of galaxy evolution
in dense environments. It may also have a significant impact on the evolution
of the orbital trajectories. As discussed in the Introduction, pre-processing is naturally
expected in the $\Lambda$CDM cosmology -- or in {\em any} hierarchical model -- due to the bottom-up assembly of dark
matter haloes. This motivates robustly quantifying the effect.

There are at least three ways in which a dark matter halo (or galaxy) may be pre-processed before being accreted
by a primary halo. It may be a backsplash halo of the primary itself, and is again infalling. Having been
previously accreted, these seemingly isolated systems have already passed through the much denser environment of the primary,
potentially falling victim to tidal or ram pressure stripping. Similarly, it may be a subhalo or (secondary)
backsplash halo of a {\em infalling} system. Cosmic web stripping is another possibility: large-scale
cosmic filaments of gas and dark matter that feed primary haloes trace out large cosmological volumes, and can have
large relative velocities with respect nearby haloes. Although of little importance for their dark matter content,
haloes that cross path with these filaments may sustain losses of (gaseous) baryons as a result of
ram pressure stripping \citep{Ale2013}.

As discussed above, roughly 59 per cent of associated subhaloes are currently beyond the virial radius of their
host despite being bound to it. This adds a level of ambiguity to what is meant by ``group accretion'' as
any infalling halo may also have its own population of secondary subhaloes and backsplash haloes.
Because of this ambiguity, we identify the members of accreted groups in two ways: First, as the {\em full} population of
(sub)haloes that were at any point prior to accretion by the primary identified as a secondary subhalo of a
more massive infalling system (these systems have been ``pre-processed'' by secondary haloes but are not necessarily
secondary subhaloes at the present time). And second, as the population of infalling systems that were identified as
secondary subhaloes within a narrow radial window -- which we initially take to be $(1-1.2)\times r_{200}$ -- prior to their
accretion by the host. The latter definition of group accretion is stricter, and demands that members of accreted
groups be secondary subhaloes near the point of infall.

\begin{figure}
  \centering
  \includegraphics[width=\linewidth]{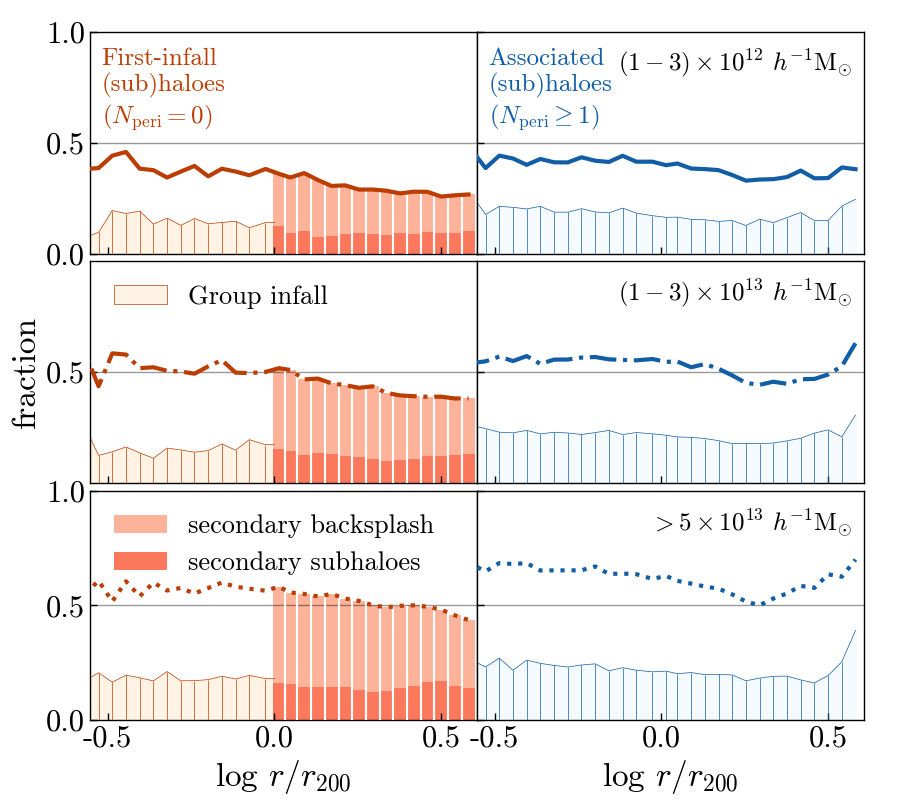}
  \caption{The relative importance of pre-processing for first-infall haloes and subhaloes (left panels)
    and for orbital subhaloes and backsplash haloes (right panels) as a function of separation from primary
    haloes of different virial mass (top to bottom). Thick lines of different line-style show the fractions
    of (sub)haloes that were identified as substructures of a more massive system
    (excluding the primary halo) prior to infall or $z=0$, which ever comes first. First-infall {\em haloes}
    ($r_{\rm sub}>r_{200}$, left panels) are divided into secondary
    subhaloes (i.e. current subhaloes of secondary haloes; dark shaded region) and secondary backsplash haloes
    (light shaded region). First-infall {\em subhaloes} ($r_{\rm sub}\leq r_{200}$, left panels) are divided into
    those accreted in isolation (white regions) or as part of a larger group (light shaded region; see
    text for a definition of how group-infall events are classified).
    Similarly, the right-hand panels show the fraction of orbital subhaloes ($r_{\rm sub}\leq r_{200}$) and
    backsplash haloes ($r_{\rm sub}>r_{200}$) that were accreted in groups (shaded regions).}
  \label{fig3}
\end{figure}

Figure~\ref{fig3} summarises the relative fraction (thick lines) of pre-processed haloes and subhaloes as a function of
distance from their primary, showing separately those on first-infall (i.e. ${\rm N_{peri}=0}$, left) and the combination of
orbital subhaloes and backsplash haloes (${\rm N_{peri}\geq 1}$, right). As in several previous figures,
results are shown for three separate primary mass bins, with mass increasing from top to bottom.

A large fraction of first-infall subhaloes ($r_{\rm sub}\leq r_{200}$, left panels) 
were substructures of more massive systems prior to
their accretion by the primary -- typically 40 to 55 per cent, depending on the primary's mass.
Of these pre-processed haloes, roughly 31 to 38 per cent (depending on primary mass) were 
satellites {\em at the time of accretion} (i.e. were identified as secondary subhaloes at their
point of entry into the primary; light shaded bars for $r\leq r_{200}$); $\approx 29$ to 32 per cent are
{\em current} secondary subhaloes of infalling haloes (dark shaded bars for $r > r_{200}$).

The right-hand panels of Figure~\ref{fig3} show the fraction of all orbital subhaloes and backsplash haloes
at a given radial separation from their host that were pre-processed by a secondary halo prior to first infall (thick lines),
or were accreted as part of a group (shaded regions). Pre-processing and group accretion
are approximately equally important for systems already accreted (right panels), and for those currently
infalling for the first time (left panels).
These results depend only weakly on the mass of the primary, and on present-day radial separation.

\begin{figure}
  \centering
  \includegraphics[width=0.9\linewidth]{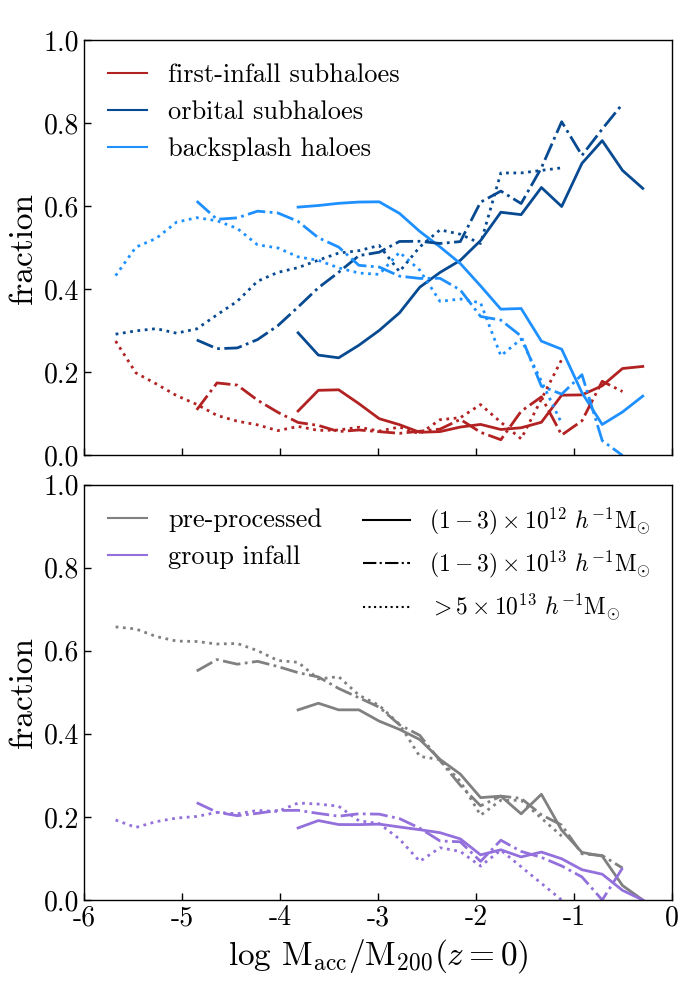}
  \caption{The fraction of haloes and subhaloes in various orbit categories as a function of their mass
    at the time of accretion, normalized by the present-day mass of their host halo. Upper panels show separately
    infalling subhaloes (i.e. ${\rm N_{peri}}=0$ and $r_{\rm sub}\leq r_{200}$; red curves), orbital
    subhaloes (${\rm N_{peri}}\geq 1$ and $r_{\rm sub}\leq r_{200}$; dark blue curves) and
    backsplash haloes (light blue); lower panels show the fractions of all haloes that were pre-processed
    prior to infall (grey curves), or were accreted as part of a larger group of haloes (purple
    lines). Different lines correspond to different host mass bins, as indicated.}
  \label{fig:grpinfallmass}
\end{figure}

In the upper panel of Figure~\ref{fig:grpinfallmass} we plot the fraction of (sub)haloes of various orbital types as a function
of their mass at accretion time, normalized by the present-day mass of their primary host halo. We use different line
styles for different primary mass bins, as in previous figures. Note the strong mass-dependence of the orbital subhalo and
backsplash halo fractions, a result already hinted at in Figure~\ref{apoevol}. The mass dependence arises because the
most massive associated subhaloes (${\rm M_{acc}}\simgt 0.01\times {\rm M_{200}}$) tend to rapidly
lose orbital energy due to dynamical friction after infall, and often remain orbital subhaloes until the present day.

As discussed in \citet{Ludlow2009}, and as evident in Figure~\ref{fig:grpinfallmass}, low-mass members of accreted groups tend to be the ones ``ejected'' to
higher energy orbits. This is because low-mass systems are more likely to be loosely bound to the group, and therefore to occupy
large-amplitude orbits that are able to capture orbital energy when in phase with the orbit of the group within
the primary. As a result, backsplash haloes tend to dominate the population of associated subhaloes below a characteristic
mass that depends on the mass of the host. 

Low-mass haloes are also more likely to be pre-processed than massive ones, as shown in the lower panel of Figure~\ref{fig:grpinfallmass}.
For example, roughly half of all present-day subhaloes whose mass at accretion time is of order
${\rm M_{acc}}\approx 10^{-3}\,{\rm M_{200}}$ were previously substructures of another (secondary) halo
Even the most massive accreted systems are not entirely immune to the effects of pre-processing: for example, roughly 15 per cent
of those with ${\rm M_{acc}/M_{200}}\approx 0.1$ had been identified as subhaloes of other, more-massive systems prior to infall,
and of order 5 per cent were substructures {\em at the time} of accretion.

\subsection{Impact of pre-processing and subhalo-subhalo encounters on the evolution of orbital trajectories}
\label{ssec:prepdynamics}

\begin{figure*}
  \centering
  \includegraphics[width=0.85\linewidth]{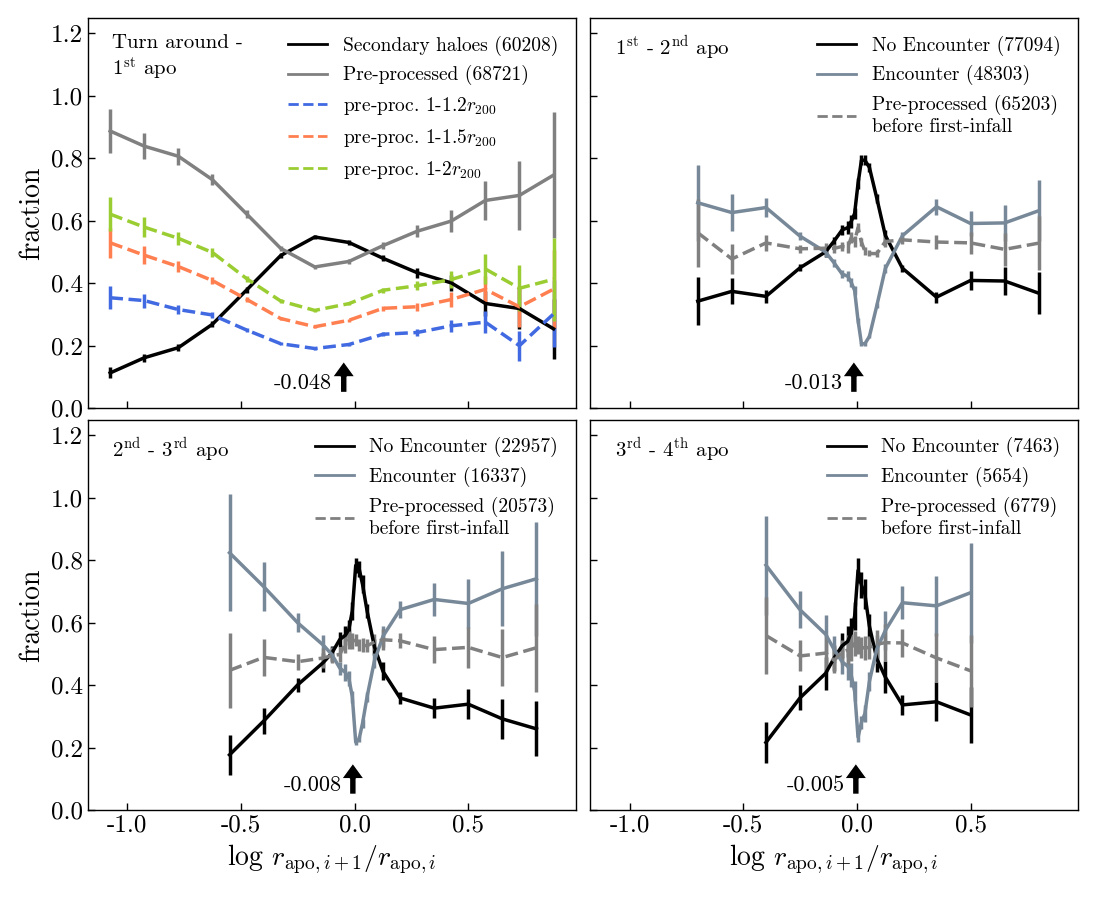}
  \caption{The fraction of associated subhaloes in bins of $\log r_{{\rm apo},i+1}/r_{{\rm apo},i}$, which
    quantifies the ratio of successive apocentric distances. As with Figure~\ref{apoevol}, different panels
    show less- or more-evolved orbits. The top-left panel, for example, corresponds to the ratio of
    the first apocentre ({\em after} infall) to the turnaround radius; the lower right panel to the ratio of the
    third-to-fourth apocentre. In the upper-left panel we divide subhaloes into two groups: one, shown using
    a solid black line, corresponds to pristine haloes (i.e. haloes that were never identified as substructures
    prior to infall), and another (grey line) showing those that were pre-processed before infall. Different
    coloured lines correspond to systems that were identified as secondary subhaloes within increasing radial apertures
    just prior to first accretion by the host (e.g. the blue line for systems that were tagged as secondary subhaloes
    between 1 to $1.2\times r_{200}$; the green lines between 1 to $2\times r_{200}$). Grey dashed lines in all other panels also correspond to
    haloes that were pre-processed prior to {\em first} infall, and solid grey lines to those that experienced a penetrating encounter with
    another subhalo between apocentre $i$ and $i+1$; solid black lines correspond to
    systems that {\em did not} encountered another subhalo along their orbital paths between the two relevant apocentres. 
    Note that the extrema of these distributions are primarily associated with subhaloes that have interacted with others along
    their orbit paths.}
  \label{fig:grpinfall}
\end{figure*}

We showed above (Figure~\ref{apoevol}) that successive apocentres of associated subhaloes evolve roughly 
as expected from simple spherical infall models, albeit with considerable variation among individual
systems. The orbital trajectories of a few carefully-selected subhaloes, shown in Figure~\ref{apoevolex}, suggest
that the accretion of dynamically-associated groups of haloes, or close encounters between subhaloes,
can lead to an exchange of orbital energy resulting in rather abrupt changes
to their orbital apocentres. Do these particular examples encompass the majority of scenarios that may lead to deviant
evolution of orbital apocentres? We explore this possibility in Figure~\ref{fig:grpinfall}.

Consider first the upper-left panel, where we plot the relative fraction of associated
subhaloes in bins of $r_{{\rm apo},1}/r_{\rm ta}$ that were either a) pre-processed prior to accretion
(grey line), b) accreted as part of a group of haloes (coloured lines), or c) as 
secondary haloes that were {\em never} identified as substructures of any other halo (solid black line).
The differences are striking. For secondary haloes (c) the ratio of first apocentre to turnaround radius peaks at around
$r_{{\rm apo},1}/r_{\rm ta}\approx 0.81$, comparable to analytic expectations.
Very few of the outlier subhaloes are among this category. Instead, the outliers
are primarily {\em pre-processed} haloes, i.e. haloes and subhaloes that are (or were) substructures of more
massive systems prior to being accreted by the host. Indeed, pre-processed haloes account for more than
two-thirds of systems with $r_{{\rm apo},1}/r_{\rm ta}\simgt 3$, and more than 80 per cent of those
with $r_{{\rm apo},1}/r_{\rm ta}\simlt 0.25$.

Many of these pre-processed haloes are members of loosely-bound groups
that are tidally disrupted near first pericentre, giving rise to the unorthodox orbits seen in
Figure~\ref{apoevol} (this includes systems that both gain {\em and} loose orbital energy). This
effect is portrayed more clearly by the coloured lines, which correspond to 
sub-samples of associated subhaloes that were pre-processed (i.e. identified as secondary subhaloes)
within increasing radial apertures from the host's centre prior to infall. The blue curve, for example, corresponds to
systems that were formally accreted in a group (according to the definition of group accretion
in section~\ref{ssec:preprocessing}, i.e. were pre-processed within a spherical aperture between 1 and $1.2\times r_{200}$
prior to infall); these systems account for the majority of the most extreme cases.
Orange and green curves, respectively, correspond to systems that were pre-processed within 1.5
and $2\times r_{200}$ when on first approach. Increasing the aperture within which dynamically-associated
infalling groups are identified clearly accounts for an increasing number of unexpectedly
large and small first apocentres.

But what about subsequent orbits? The fact that many members of accreted groups are likely tidally dispersed
on first pericentric passage suggests that whether a halo was pre-processed or not is unlikely to determine
whether energy is lost or gained thereafter. This expectation is borne out by our
simulations. The dashed grey lines in the remaining three panels, for example, show the ratio of
subsequent apocentres (first-to-second to third-to-fourth) for systems that were pre-processed prior to {\em first}
infall. In all cases this distribution remains approximately flat: evidently, pre-processing greatly
affects the extreema of {\em first} apocentres, but not subsequent ones. 

As discussed in Section~\ref{ssec:apoevol} \citep[see also][]{Tormen1998b,vandenBosch2017},
penetrating encounters between subhaloes are common, typically affecting 60 per cent of all
associated subhaloes at some point along their orbital paths. The majority of these encounters occur near
pericentre -- i.e. as close to the potential minimum of the host halo as their orbit allows -- where subhalo
number densities are highest, and therefore may constitute $\geq 3$-body interactions. Such 3-body interactions
allow for a rapid redistribution of orbital energy between the interacting bodies and, as with infalling
groups of haloes, can readily perturb orbits to higher or lower energies.

The remaining panels in Figure~\ref{fig:grpinfall} show that penetrating encounters indeed affect the
evolution of subsequent orbital apocentres. Each panel plots the relative fraction of haloes, in bins
of $r_{{\rm apo},i+1}/r_{{\rm apo},i}$, that either experienced a penetrating encounter along
that particular segment of their orbit (grey lines) or did not (black lines). In all cases, the
extreema tend to be associated with those subhaloes that have experienced a penetrating encounter
{\em between} $r_{{\rm apo},i}$ and $r_{{\rm apo},i+1}$; those that have not tend to follow
more closely the evolution expected from simple spherical infall models (the arrows in each panel
mark the expectations from Bertschinger's model). 

\subsection{The structure of first-infall and associated haloes and subhaloes}
\label{ssec:struct}

It is well-established that the structure of dark matter haloes is largely determined by their
collapse time \citep[e.g.][]{Bullock2001,Neto2007,Ludlow2013,Ludlow2014,Correa2015c},
but also depends on environment \citep{AvilaReese2005}, as well as on the various
interaction histories haloes may have had with their neighbours \citep{Li2013,Wang2020}. 
Correlations between the structure of haloes and their varied assembly histories are often difficult
to disentangle, but give rise to important secondary phenomena. One, known as assembly bias,
quantifies the clustering strength of haloes as a function of their concentration or formation time
\citep[e.g.][]{Gao2005}: at fixed mass, older haloes cluster more strongly than younger ones, highly concentrated
haloes more than less concentrated ones. A variety of interpretations have been put forth for the origin
of the age-dependence of halo clustering, although it appears that no single process can explain it
entirely \citep[see, e.g.,][for a recent discussion]{Mansfield2020}. For example, backsplash haloes have
undergone tidal stripping by their hosts and appear {\em older} than field haloes of similar mass; they
are also, by selection, more clustered due to their proximity to more massive hosts. Nevertheless, the
assembly bias remains even after backsplash haloes have been accounted for \citep{Wang2009}, suggesting
that other dynamical processes are at work. Possibilities include the suppression of halo growth in dense
environments due to tidal forces from large-scale structure \citep[e.g.][]{Hahn2009,Hearin2016}, or large-scale
tidal anisotropies \citep{Paranjape2018}. 

\begin{figure*}
  \centering
  \includegraphics[width=0.85\linewidth]{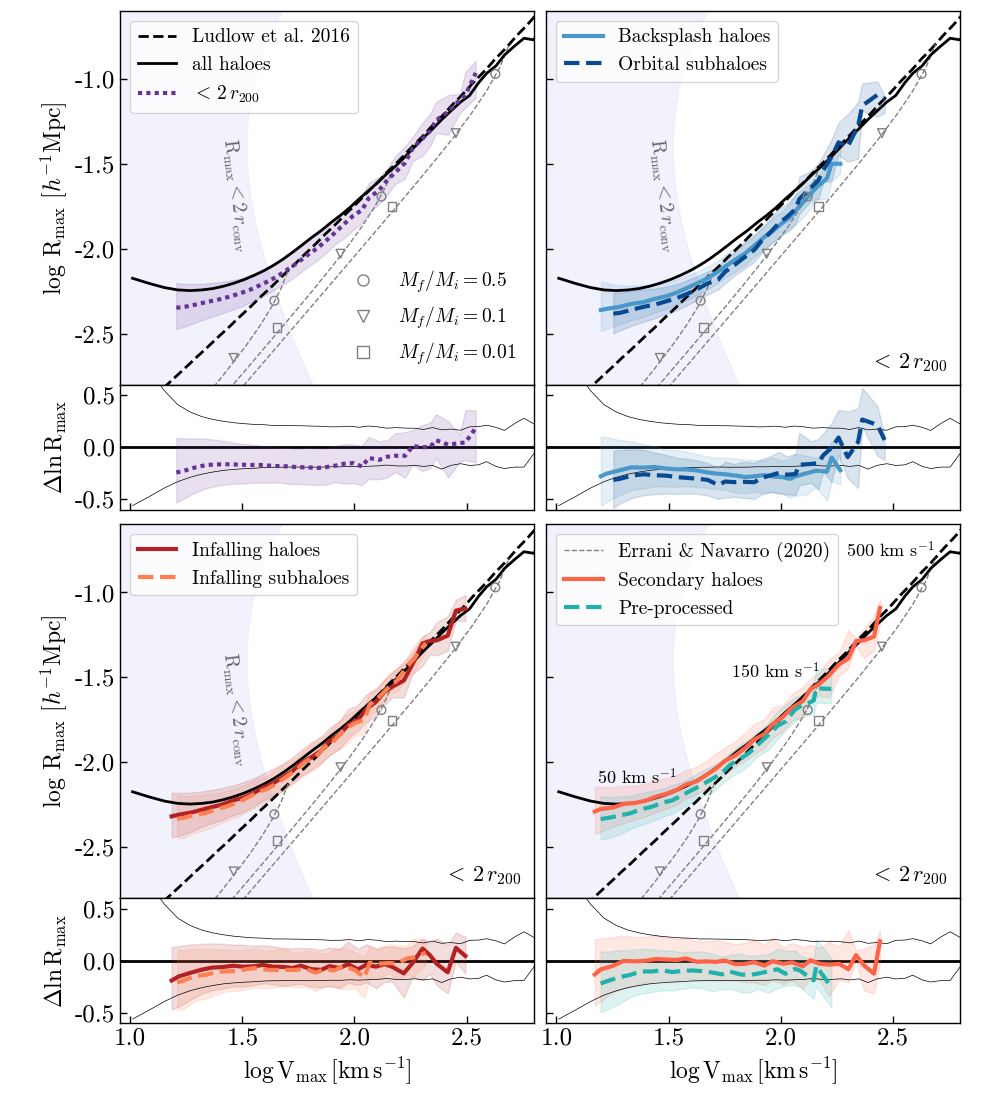}
  \caption{The ${\rm R_{max}-V_{max}}$ relations for the various categories of haloes and subhaloes studied
    in this paper. The upper-left panel compiles results for all haloes and subhaloes that lie within 2 virial
    radii from their respective hosts (purple dotted line); the upper-right panel shows separately the median
    relations for all {\em orbital} subhaloes (i.e. ${\rm N_{peri}\geq 1}$ and $r_{\rm sub}\leq r_{200}$; dashed blue lines)
    and backsplash haloes (i.e. ${\rm N_{peri}\geq 1}$ and $r_{\rm sub} > r_{200}$; solid blue lines); the lower-left
    panel distinguishes infalling secondary haloes (solid red line) and secondary subhaloes (dashed orange
    line); the lower-right panel distinguishes pre-processed haloes (green dashed line) from isolated ones (solid
    orange). Note that, in all cases, we restrict the subhalo samples to those that lie within $2\times r_{200}$
    from the primary. The black solid line, repeated in all panels for comparison, shows the ${\rm R_{max}-V_{max}}$ relation for
    {\em all} field haloes in the simulation volume (no isolation criteria are imposed). The dashed grey lines
    show the ${\rm R_{max}-V_{max}}$ relation predicted by the analytic model of \citet{Ludlow2016}. The 
    curve delineating the open and shaded regions corresponds to the ${\rm R_{max}-V_{max}}$ relation for
    a halo whose mass is determined by the constraint ${\rm R_{max}}=2\times r_{\rm conv}$, where
    $r_{\rm conv}$ is the convergence radius defined by \citet{Ludlow2019a}. Thin dashed lines show
      several ``tidal tracks'' predicted by the model of \citet{Errani2020} for haloes with ${\rm V_{max}=50}$,
      150 and 500 km/s at infall; the symbols along each of these curves indicate the surviving bound mass fraction.}
  \label{fig:vrmax}
\end{figure*}

In Figure~\ref{fig:vrmax} we quantify how the orbital histories of haloes and subhaloes affect
their concentrations -- a useful proxy for formation time -- which we characterize using the magnitude and
location of the maximum circular
velocity, ${\rm V_{max}}$ and ${\rm R_{max}}$, respectively. This is a robust proxy for concentration
for a couple of reasons: first, it can be estimated non-parametrically, and is therefore free from subtle
biases that may be introduced when fitting to some suitably-smooth profile, such as NFW; second, while tidal
stripping may substantially reduce the dark matter mass of a halo, its impact on ${\rm V_{max}}$ and ${\rm R_{max}}$ is
much less dramatic \citep[e.g.][]{Gao2012}.

The different panels of Figure~\ref{fig:vrmax} compare the ${\rm V_{max}} - {\rm R_{max}}$ relations for the various samples
of (sub)haloes to that of the {\em entire} population of haloes in the simulation (shown as a solid black line and repeated in
all panels for comparison). Not surprisingly, the full population of (sub)haloes that lie within 2 virial
radii of their hosts (dotted purple line, upper-left panel) are more concentrated than the average field halo,
typically by about $\approx $15 per cent (recall that roughly half of these are ``associated'' subhaloes
and half are on first infall).

The upper-right and lower-left panels distinguish orbital (sub)haloes from those on first infall, respectively,
and in each case the two samples have been subdivided into those inside and outside of $r_{200}$.
Orbital subhaloes and backsplash haloes (solid and dashed blue lines in the upper-right panel) typically have
concentrations that are of order 22 per cent higher than field haloes, except possibly for the highest
${\rm V_{max}}$. Interestingly, infalling (sub)haloes have similar concentrations to field haloes -- smaller
by only a few per cent -- regardless of their radial separation from the host (the dashed orange line, for
example, corresponds to {\em subhaloes} that are on first infall, whereas the solid red line corresponds to infalling
haloes at $r > r_{200}$).

The lower-right panel divides infalling haloes further, into ``pristine'' secondary haloes, i.e.
those that have never been satellites of more massive systems (solid orange line), and pre-processed haloes
that have (dashed green line; these are secondary backsplash haloes and secondary subhaloes). Secondary haloes on first infall
have concentrations that are indistinguishable from those of field haloes, even when the former are restricted to radial
separations $r\leq 2\times r_{200}$; pre-processed haloes are, on average, $\approx 10$ per cent more
concentrated. These results agree with the conclusions of \citet{Li2013}, who claim that tidal interactions between
backsplash haloes and their former hosts are the primary mechanisms by which the concentrations of haloes in
the environs of more massive systems are modified. 

The systematic differences in the ${\rm V_{max}}-{\rm R_{max}}$ relations for haloes of different orbital type
  likely originate from their different tidal mass loss histories. We show this using ``tidal tracks'' (predicted by the
  model of \citealt{Errani2020}; see also \citealt{Hayashi2003,Pen2010,Green2019}), which describe
  the evolutionary paths followed by haloes in the ${\rm V_{max}}-{\rm R_{max}}$ plane as a function of the mass tidally
  stripped since infall. Curves are shown for three initial values of ${\rm V_{max}}$ (50, 150 and 500 km/s) and
  start from the dashed line \citep[the model of][]{Ludlow2016}. Symbols along each line mark remnant mass fractions
  corresponding to 0.5 (circle), 0.1 (triangle) and 0.01 (square). 

\section{Summary}
\label{sec:summary}

We used high-resolution, cosmological dark matter-only simulations to classify the orbital histories of
haloes and subhaloes surrounding isolated hosts. Our analysis targets hosts that span the mass range
$10^{12}\leq {\rm M_{200}}/[h^{-1} {\rm M}_\odot]\leq 3.4\times10^{14}$ ($8.6\times10^4\leq {\rm N_{200}}\leq 2.9\times10^7$; 2309 in total).
As in previous work, we characterized the full population of haloes that ever crossed the
virial radius of their host (so-called ``associated'' subhaloes) but considered, in addition, the
remaining population that lie within four virial radii of the host's centre-of-potential. We carefully
tracked the orbital trajectories for each of these haloes and subhaloes, tallying all apsis
points in order to identify an unambiguous sample of ``first-infall'' haloes and subhaloes -- i.e.
those have that have not yet crossed pericentre on their orbit about the host -- as well as ``orbital''
haloes and subhaloes -- i.e. those that have completed at least one pericentric passage. By doing so, we were able
to easily distinguish both haloes {\em and} subhaloes that are infalling for the first time
from those that occupy more evolved orbits. Orbital systems are either ``typical'' substructure
(i.e. have $r_{\rm sub}\leq r_{200}$ at $z=0$), backsplash haloes ($r_{\rm sub} > r_{200}$ at $z=0$ but
$r_{\rm sub} < r_{200}$ in the past), or, less often, haloes whose
pericentres occur {\em outside} their host's virial radius (referred to as ``orbital'' haloes). The
variety of possible orbital trajectories considered in this paper are defined in Section~\ref{ssec:orbits} (see also Figure~\ref{fig0});
their distribution in radial velocity-distance phase-space is shown in Figure~\ref{fig1}.

In agreement with previous work \citep[e.g.][]{Diemand2004b,Reed2005,Gao2004,Han2016}, we find that the spatial 
distribution of associated subhaloes is substantially less concentrated towards the halo centre than the dark matter
Their radial number density profiles, for example, can be well approximated by an Einasto profile with shape parameter
$\alpha\approx 1.2$ (the concentration parameter is $c\approx 2.6$),
implying a substantially shallower ``cusp'' than the traditional NFW profile. However,
the number density of subhaloes {\em does not} approach a constant-density core, but rather increases
(slowly) all the way to the halo centre (Figure~\ref{fig2}). We find that half of all associated subhaloes
are located within $\approx 1.24 \times r_{200}$ from the centre of their hosts, and $\approx 95$ per cent are located within
$2.6 \times r_{200}$; associated systems make up 77 and 63 percent of {\em all} (sub)haloes within these
radii, respectively. 

We find that a large fraction of classically-defined (surviving) substructures (i.e. those with $r_{\rm sub}\leq r_{200}$)
were accreted very recently: roughly 15 per cent in the past half-crossing time,
$t_{\rm cross}/2=r_{200}/{\rm V}_{200}\approx 1.72\,{\rm Gyr}$ (90 per cent of which are first-infall subhaloes),
and 21 per cent in the interval
$t_{\rm cross}/2\leq t_{\rm acc}\leq t_{\rm cross}$ (primarily orbital subhaloes still approaching their first
apocentre). Backsplash haloes of were primarily accreted at lookback times
exceeding $t_{\rm cross}$, which is the typical timescale required for recently-accreted material to first
exit the virial boundary of its host halo.
This suggests that many orbital subhaloes are recent arrivals: roughly half have crossed pericentre
only once, and roughly one-third have yet to reach their first apocentre (after turnaround). Only $\approx 56$ per cent
of {\em all} subhaloes within $r_{200}$ have completed at least one apocentric passage since infall (Figure~\ref{fig:zacc}).

Splashback radii are therefore not directly measurable from orbital tracks for many associated subhaloes, but
may nevertheless be estimated provided
their orbital energy and angular momentum is known.
The ratio of consecutive apocentres (including both measured and predicted ones) for associated subhaloes
with mass ratios ${\rm M_{sub}/M_{200}}\simlt 0.01$ approximately follow the pattern expected 
from simple self-similar infall models \citep[e.g.][]{Bertschinger1985}, although
with slightly greater loss of orbital energy after first pericentre. Bertschinger's model, for
example, predicts that the first apocentre of a typical orbit should be of order 90 per cent of
its turn around radius; our simulations suggest a median value of
$r_{{\rm apo},1}/r_{\rm ta}\approx 0.8$ (or a mean value of 0.87) for the lowest-mass subhaloes.
Nevertheless, as expected
from analytic models, the loss of orbital energy is most pronounced after first pericentric
passage, after which the orbits of low-mass subhaloes typically reach a stationary
state, and the ratio of successive apocentres approaches unity.
These results {\em do not} apply to the orbits of the most massive subhaloes (${\rm M_{sub}/M_{200}}\simgt 0.01$),
which are repeatedly degraded due to dynamical friction (See Figure~\ref{apoevol}.)

Although these trends apply {\em on average}, the orbits of individual subhaloes exhibit a large
variation between successive apocentres. For example, for hosts with ${\rm M_{200}}\approx 10^{12} h^{-1}{\rm M_\odot}$
($5\times 10^{13}h^{-1}{\rm M_\odot}$), roughly 26 (25) per cent of associated subhaloes have first apocentres
that exceed their nominal turnaround radius, whereas $\approx 10$ (15) per cent have $r_{{\rm apo},1}/r_{\rm ta}\simlt 0.5$.
These outliers -- common in cosmological haloes -- deviate significantly from simple, analytic predictions.
The discrepancy between orbits in simulations and those expected from analytic models appear to arise due to
high-order interactions between subhaloes and the hosts in which they orbit. On first approach,
the accretion of loosely-bound groups of haloes can account for the vast majority of outlier {\em first} apocentres (roughly
two-thirds of those with $r_{{\rm apo},1}/r_{\rm ta}\simgt 3.3$, and 80 per cent of those with
$r_{{\rm apo},1}/r_{\rm ta}\simlt 0.25$ were pre-processed prior to accretion). Rapid changes in orbital 
energy on subsequent orbits is mainly a result of penetrating encounters between subhaloes (see Figure~\ref{fig:grpinfall}). 

The orbital histories of (sub)haloes also affects their internal structure, results
which we summarize in Figure~\ref{fig:vrmax}. Infalling secondary 
haloes (i.e. those that have {\em never} been substructures of more massive haloes) follow a ${\rm V_{max}-R_{max}}$
relation that is similar to the entire population of field haloes.
Haloes that have been pre-processed, but not yet accreted by the host, are systematically more concentrated.
Orbital subhaloes and backsplash haloes -- which, by definition, have completed at least one pericentric passage about their host
haloes -- exhibit, relative to isolated field haloes, the most discrepant concentrations. These results imply that
tidal encounters between associated subhaloes and their hosts are the primary mechanism
by which the structural scaling relations of dark matter haloes are modified in dense environments. 

\section*{Acknowledgements}

We are grateful to our referee, Frank van den Bosch, for a useful report,
which led to a number of improvements to our paper.
We acknowledge various public {\textsc{python}} packages that have 
benefited our study: {\textsc{scipy}} \citep{scipy}, {\textsc{numpy}} \citep{numpy},
{\textsc{matplotlib}} \citep{matplotlib} and {\textsc{ipython}} \citep{ipython}.
ADL acknowledges financial support from the Australian Research Council through their
Future Fellowship scheme (project number FT160100250). Parts of this research were conducted
by the Australian Research Council Centre of Excellence for All Sky Astrophysics in 3
Dimensions (ASTRO 3D), through project number CE170100013. This research/project was undertaken
with the assistance of resources and services from the National Computational Infrastructure
(NCI), which is supported by the Australian Government, and supported by
resources provided by the Pawsey Supercomputing Centre with funding from the Australian
Government and the Government of Western Australia.

\bibliographystyle{mn2e}
\bibliography{paper}

\end{document}